\documentclass[twocolumn,twocolappendix]{aastex631} 

\usepackage{amsmath,amssymb,amsfonts,bm}
\usepackage{epstopdf}
\usepackage{epsfig}
\usepackage{natbib,caption2}
\usepackage{graphicx}   
\usepackage{float}
\usepackage{longtable}
\usepackage{graphics}
\usepackage{hyperref}
\usepackage{color}
\usepackage{calc}
\usepackage{threeparttable}

\newcommand{\magsqarc}{mag\,arcsec\,$^{-2}$}
\newcommand{\msun}{M$_\odot$}
\newcommand \beq{\begin{equation}}
\newcommand \eeq{\end{equation}}
\newcommand \bey{\begin{eqnarray}}
\newcommand \eey{\end{eqnarray}}

\newcommand \Lsun{L$_\odot$}

\newcommand{\gsim}{\lower.5ex\hbox{$\; \buildrel > \over \sim \;$}}
\newcommand{\lsim}{\lower.5ex\hbox{$\; \buildrel < \over \sim \;$}}

\accepted{ApJS}

\shortauthors{Tao et al.}

\begin{document}

\title{The First Systematic Survey of Stellar Halos in High-Inclination Galaxies Reveals Unusually Quiescent Merger Histories of Nearby Galaxies}

\author[0000-0002-8420-6246]{Bojun Tao}
\email{hzhang18@ustc.edu.cn}
\affiliation{Department of Astronomy, University of Science and Technology of China, Hefei, Anhui 230026, China}
\affiliation{School of Astronomy and Space Science, University of Science and Technology of China, Hefei, Anhui 230026, China}

\author[0000-0003-1632-2541]{Hong-Xin Zhang}
\altaffiliation{Corresponding author}
\affiliation{Department of Astronomy, University of Science and Technology of China, Hefei, Anhui 230026, China}
\affiliation{School of Astronomy and Space Science, University of Science and Technology of China, Hefei, Anhui 230026, China}

\author[0000-0002-5762-7571]{Wenting Wang}
\affiliation{Department of Astronomy, Shanghai Jiao Tong University, Shanghai 200240, China}
\affiliation{State Key Laboratory of Dark Matter Physics, Key Laboratory for Particle Astrophysics and Cosmology (MOE), \& Shanghai Key Laboratory for Particle Physics and Cosmology, Shanghai Jiao Tong University, Shanghai 200240, China}

\author[0000-0003-1588-9394]{Enci Wang}
\affiliation{Department of Astronomy, University of Science and Technology of China, Hefei, Anhui 230026, China}
\affiliation{School of Astronomy and Space Science, University of Science and Technology of China, Hefei, Anhui 230026, China}

\author[0000-0002-4742-8800]{Guangwen Chen}
\affiliation{Sub-department of Astrophysics, University of Oxford, Keble Road, Oxford, OX1 3RH, UK}
\affiliation{Instituto de Astrof\'isica de Canarias, calle Vía L\'actea s/n, E-38205 La Laguna, Tenerife, Spain}
\affiliation{Departamento de Astrof\'isica, Universidad de La Laguna, Avenida Astrof\'isico Francisco S\'anchez s/n, E-38206 La Laguna, Spain}

\author{Huiyuan Wang}
\affiliation{Department of Astronomy, University of Science and Technology of China, Hefei, Anhui 230026, China}
\affiliation{School of Astronomy and Space Science, University of Science and Technology of China, Hefei, Anhui 230026, China}

\author{Lijun Chen}
\affiliation{Department of Astronomy, University of Science and Technology of China, Hefei, Anhui 230026, China}
\affiliation{School of Astronomy and Space Science, University of Science and Technology of China, Hefei, Anhui 230026, China}

\author[0000-0002-4382-1090]{Qian-Hui Chen}
\affiliation{Research School of Astronomy \& Astrophysics, Australian National University, Canberra, Australia, 2611}
\affiliation{ARC Centre of Excellence for All Sky Astrophysics in 3 Dimensions (ASTRO 3D), Australia}

\author[0000-0003-1385-7591]{Song Huang}
\affiliation{Department of Astronomy, Tsinghua University, Beijing 100084, China}

\author[0000-0002-7660-2273]{Xu Kong}
\affiliation{Department of Astronomy, University of Science and Technology of China, Hefei, Anhui 230026, China}
\affiliation{School of Astronomy and Space Science, University of Science and Technology of China, Hefei, Anhui 230026, China}

\author{Yu Rong}
\affiliation{Department of Astronomy, University of Science and Technology of China, Hefei, Anhui 230026, China}
\affiliation{School of Astronomy and Space Science, University of Science and Technology of China, Hefei, Anhui 230026, China}

\begin{abstract}

Stellar halos are the only major stellar component of disk galaxies that lack systematic observational characterization, yet they encode critical information about galaxy merger histories. We present the first systematic census of stellar halos in a large, flux-limited sample of 169 high-inclination central galaxies with stellar masses $7.3 \leq \log M_{\star}/M_{\odot} \leq 11.0$ and redshift $z < 0.1$, using HSC-SSP Deep optical images. Stellar halos are detected in 93 galaxies, primarily through their low isophotal ellipticities in the outskirts, improving upon conventional methods of stellar halo identification. The halo detection rate reaches $\sim$ 50\% at $\log M_{\star}/M_{\odot} > 9.9$ and $\gtrsim$ 70\% for Milky Way (MW)-mass galaxies. We derive halo surface brightness profiles, colors, and masses, finding that stellar halos generally follow power-law radial profiles. Higher-mass galaxies, on average, exhibit smaller power-law indices and larger halo mass fractions, indicating more extended halos and more active merger histories. A significant stellar halo color–mass correlation, driven mainly by the mass–metallicity relation, suggests dominance by a few massive accretion events. MW-mass galaxies have a median stellar halo fraction of $10\%\pm5\%$. Among nearby galaxies with halo measurements within 25 Mpc, two thirds (including the MW) lie below the mean stellar halo fraction–galaxy mass relation. Overall, the nearby galaxies show a median halo deficit of $\sim$ 0.3 dex, implying unusually quiescent merger histories. We show that this deficit follows a broader trend in which typical halo fractions increase with heliocentric distance, tracking the gradual rise in matter density toward the cosmic average by $z$ $\lesssim$ 0.07.

\end{abstract}


\section{Introduction}

According to the standard $\Lambda$ Cold Dark Matter ($\Lambda$CDM) cosmological model, galaxies form near the centers of dark matter halos and grow through accretion along the cosmic web and through mergers with other galaxies \citep{White_1978, Moore_1999}. In this hierarchical framework, galaxy interactions and mergers play an important role in driving the evolution of galaxies, especially during the early cosmic epoch. Galaxies that are fully or partially disrupted during mergers contribute to the buildup of stellar halos in the merger remnants \citep{Bullock_2005, Cooper_2010}.

Cosmological simulations suggest that stellar halos—characterized by relatively long dynamical timescales \citep{Bullock_2005} and high ex-situ stellar fractions \citep{Font_2011}—act as relics of galaxies’ hierarchical assembly histories. State-of-the-art hydrodynamical simulations \citep{Oser_2010, Cooper_2010, Wang_2011, Lackner_2012, Cooper_2013, Pillepich_2014, Cooper_2015, Rodriguez-Gomez_2016, DSouza_2018, Karademir_2019, Rey_2022, Keller_2022, Shi_2022, Genina_2023, Angeloudi_2024, Grimozzi_2024, Cooper_2025}, such as the Illustris \citep{Rodriguez-Gomez_2016} and IllustrisTNG \citep{Shi_2022, Angeloudi_2024} simulations, showed that the fraction of stars formed ex-situ (and thus the stellar halo mass fraction) strongly depends on galaxy stellar mass and is generally higher in more massive and elliptical galaxies. Nevertheless, \cite{Rey_2022} demonstrate that there is a wide spread (1.5 dex) in stellar halo mass fractions due to the diversity of merging histories. Due to the steep decline of stellar-to-dark matter halo mass ratio toward the low mass end, the stellar halos of massive galaxies are expected to primarily consist of stars stripped from a small number of relatively massive satellites \citep[e.g.,][]{Genina_2023}.

Due to their proximity, the stellar halos of the Milky Way (MW) and M31 have been studied in the greatest detail. By counting individual stars and globular clusters (GCs), various halo properties have been revealed \citep{Helmi_2008}, including the dynamical state and stellar orbits \citep{Eggen_1962, Carney_1986, Chiba_2000, Yanny_2003, Carollo_2010, Belokurov_2018}, age and metallicity \citep{Hartwick_1976, Searle_1978, Harris_1979, Helmi_1999, Ibata_2001, Reitzel_2002, Roederer_2009, An_2013, Carollo_2016, Fern_2017, Youakim_2020, Viswanathan_2024}, and structural properties \citep{Woltjer_1975, Harris_1976, Preston_1991, Racine_1991, Wetterer_1996, Bell_2008, Sesar_2011, Xue_2015, Xu_2018, Chen_2023}. The stellar halos of the MW and M31 differ significantly in spatial distribution, chemical abundance, and mass \citep{Reitzel_2002, Deason_2013, Harmsen_2017}. M31’s more massive halo is thought to have formed through a major merger involving a third galaxy in the Local Group history \citep{Hammer2018}. The compact elliptical M32 is likely the remnant bulge of this disrupted galaxy \citep{DSouza_2018_NatAs, McConnachie_2018}. The revolutionary discoveries enabled by the Gaia mission revealed that, except for the very central regions, the inner and outer parts of the stellar halos of the MW are each dominantly contributed by a single accreted satellite \citep[e.g.,][]{Naidu2020}, and even the central in-situ stellar halos formed by splashed disk population are also intimately linked to past merger events \citep{Helmi_1999, Bell_2008, Helmi_2018, Mackereth_2019, Myeong_2019, Ablimit_2022, Xiang_2022, Deason2024}, reinforcing the notion that stellar halos serve as direct tracers of galaxy merger histories. 

Stellar halos can be resolved into individual stars only for galaxies within the Local Volume \citep{Mouhcine_2005, Tikhonov_2005, Seth_2007, Mouhcine_2007, Ibata_2009, Rejkuba_2009, Mouhcine_2010, Tanaka_2011, Peacock_2015, Monachesi_2016, Bell_2017, Harmsen_2017, Greggio_2018, Cohen_2020, Jang_2020, Smercina_2020, Rejkuba_2022, Smercina_2022, Gozman_2023, Harmsen_2023, Smercina_2023}. Based on resolved star counts, these studies reveal substantial diversity in stellar halo mass and metallicity, as well as a stellar halo mass-metallicity correlation, among MW-mass galaxies \citep{Bell_2017, Harmsen_2017, Smercina_2022}. While limited in scope and sample size, the resolved stellar halo studies of these very nearby galaxies provide the gold standard when exploring their unresolved counterparts in more distant galaxies.

Beyond the Local Volume, only the integrated surface brightness distribution of stellar halos can be obtained \citep{vanDokkum_2014, Trujillo_2016, Merritt_2016, Gilhuly_2022}. These studies found a wide spread in stellar halo fractions for spiral galaxies less massive than the MW, as well as a positive correlation between halo fraction and galaxy stellar mass \citep{Merritt_2016, Gilhuly_2022}. However, these studies are based on incomplete samples within a few tens of Mpc. Furthermore, this very nearby volume appears to have a significantly lower matter density than the cosmic average \citep{Keenan2013}. Consequently, the stellar halo properties and the galaxy assembly histories they imply may not be representative of the broader local universe.

Stellar halos typically exhibit surface brightness levels fainter than $\sim$28–30 \magsqarc, necessitating very deep—and thus costly—imaging observations. This imposes severe limitations on sample size and completeness, preventing such studies from yielding statistically meaningful conclusions about stellar halos as an important structural component of galaxies. To circumvent this limitation, stellar halos have also been investigated through image stacking \citep[e.g.,][]{Zibetti_2004, Tal_2011, DSouza_2014, Wang_2019, Wang_2021} or surface profile stacking \citep[e.g.,][]{Williams_2024} techniques applied to large samples of galaxies. While stacking enhances the signal-to-noise ratio and allows for the measurement of average stellar halo properties out to large galactocentric distances across galaxies of varying masses, it obscures the considerable diversity of stellar halos among individual galaxies,thereby limiting the usability of stellar halos as an indicator of galaxy assembly histories and its connection to other galaxy properties.

Existing deep imaging observations of nearby stellar halos are primarily from two  dedicated surveys: Dragonfly Nearby Galaxy Survey with the Dragonfly Telephoto Array \citep{vanDokkum_2014, Merritt_2016} and the LBT Imaging of Galaxy Haloes and Tidal Structures (LIGHTS) survey \citep{Trujillo2021, Zaritsky2024}. The two surveys push the observational boundaries of disk galaxy structures into the stellar halo regime for the first time in a systematic way, using unprecedented deep imaging (30–31\magsqarc\ in the $r$ or $V$ band) of representative galaxy samples within $\sim$30 Mpc.

Looking beyond the nearby galaxies, the deep and wide-field coverage of the Hyper Suprime-Cam Subaru Strategic Survey \citep[HSC-SSP;][]{Aihara_2018} provide the first opportunity to extend stellar halo studies to a homogeneous and statistically significant sample of galaxies in the broader local universe, without relying on image stacking. In this work, we utilize the optical images from the HSC-SSP  Deep/Ultradeep layers Public Data Release 3 \citep[PDR3;][]{Aihara_2022} to characterize the stellar halo properties of individual high-inclination central galaxies in the local universe (z $<$ 0.1).

This paper is organized as follows. Section~\ref{sec:data} describes the data and initial sample selection. Section~\ref{sec:imageprocess} details our procedures for removing background/foreground contamination and correcting for PSF scattering effects. In Section~\ref{sec:indenthalo}, we outline the methods used to extract surface brightness profiles along the minor axis, identify stellar halos, and estimate their stellar masses. Our main results are presented in Section~\ref{sec:results}, and our conclusions are summarized in Section~\ref{sec:summary}.

Throughout this paper, we adopt the first-year Planck cosmology \citep{Planck_2014}, with a Hubble constant of $H_0 = 67.3\ \mathrm{km,s^{-1},Mpc^{-1}}$, matter density $\Omega_m = 0.315$, and cosmological constant $\Omega_\Lambda = 0.685$.

\section{Data and sample} \label{sec:data}

We begin by selecting a parent spectroscopic sample of central galaxies from the SDSS NYU Value-Added Galaxy Catalogue \citep[VAGC;][]{Blanton_2005} and the Fourth Data Release of the Galaxy And Mass Assembly survey \citep[GAMA DR4;][]{Driver_2022}. The selection is limited to galaxies within the redshift range 0.001 $<$ $z$ $<$ 0.1 and located within the sky area covered by the HSC-SSP Deep/Ultradeep layers. The upper redshift limit is imposed to mitigate the cosmic dimming effect. We then retrieve and analyze the corresponding PDR3 imaging data. The remainder of this section provides further details on the data sources and sample selection.

\subsection{Selection of isolated central galaxies} \label{sec:centralgalaxy}

The parent spectroscopic sample is selected from VAGC and GAMA DR4.\ For VAGC, we use the post-redshift sample all0 as our starting point. The VAGC spans a redshift range from 0.001 to 0.4 and includes galaxies with $r$-band apparent magnitude brighter than 17.77 mag. GAMA DR4 covers a broader redshift range and reaches a deeper flux limit of 19.9 mag in the $r$ band. To ensure consistency between the two datasets, we restrict our GAMA sample to galaxies with $r$-band magnitudes brighter than 17.77 mag. The parent catalogs provide stellar masses of the galaxies, based on spectral energy distribution modeling of multi-band photometry, including that from the SDSS images. Recent studies \citep{DSouza2015, Bernardi2013} demonstrated that the standard SDSS photometry tends to underestimate total galaxy light (and thus stellar mass), missing the outer low surface brightness regions. Indeed, the SDSS photometry is systematically fainter than our measurements from the HSC-SSP images for galaxies with $\log M_{\star}/M_{\odot} < 9$. We therefore derive the stellar masses of these low-mass galaxies using the HSC-SSP $r$-band luminosity and $g-r$ color, based on the $(g-r)$--mass-to-light ratio (M/L) relation calibrated for local dwarf galaxies by \cite{Zhang2017}. We verified that adopting different mass estimation methods for low- and high-mass galaxies does not introduce systematic bias by comparing stellar masses derived from both methods for the high-mass subsample.

We select isolated central galaxies from the parent sample by largely following the methodology of \cite{Wang_2021}. In particular, an isolated central galaxy is 1) the brightest in $r$ band within $R_{\rm vir}$ and $3 \times V_{\rm vir}$ of itself and 2) outside of $R_{\rm vir}$ and $3 \times V_{\rm vir}$ of any other more massive galaxies and 3), for that from the VAGC sample, without a brighter photometric companion (see below for details), where $R_{\rm vir}$ is the virial radius of host dark matter halos and $V_{\rm vir}$ is the virial velocity. Both $R_{\rm vir}$ and $V_{\rm vir}$ are derived using the abundance matching relation from \cite{Guo_2010}.

The SDSS spectroscopic survey is affected by the ``fiber-collision'' problem, where two fibers cannot be placed closer than 55\arcsec, potentially leading to missed massive companions that lack spectroscopic redshift measurements. To mitigate this potential issue, in addition to the spectroscopic-based selection, we utilize the photometric redshift catalog Photoz2 from SDSS DR7 to exclude galaxies that have a brighter photometric companion whose redshift probability distribution has more than a 10\% chance of matching the spectroscopic redshift of the candidate central galaxy in question.

Among the above-selected isolated central galaxies, 978 are located within the HSC-SSP Deep layer footprint. We further exclude 164 galaxies that are either undergoing ongoing tidal interactions (based on visual inspection), fall within the PSF halo of bright stars (defined as regions with $g$-band $\mu_{\rm star} < 27.5$ \magsqarc around the bright star), lie near the boundary of the HSC-SSP Deep layer footprint (see below), or are significantly contaminated by nearby galaxies or artificial ghosts. Furthermore, our focus in this work is on nearly edge-on galaxies, because these galaxies offer ideal viewing angles for probing stellar halos above the disk plane. Therefore, we perform a preliminary visual inspection to exclude 565 (nearly) face-on systems. Applying all the above selection criteria, we obtain a sample of 249 isolated inclined central galaxies.

\subsection{HSC-SSP Deep survey images} \label{sec:hscimg}

The HSC-SSP Deep layer covers approximately 27 deg$^2$ in the optical and near-infrared (NIR) $g$, $r$, $i$, $z$, and $y$ broadband filters, achieving an $r$-band imaging depth of $\sim$27 mag for point sources and a median $i$-band seeing of 0.6 arcsec. The HSC-SSP Ultradeep layer, which spans 3.5 deg$^2$, is embedded within the deep layer's sky coverage and reaches about 0.6 mag deeper in imaging depth. Our analysis focuses on the $g$, $r$, and $i$ bands. The deep layer was designed to overlap with four of the most extensively observed high Galactic-latitude fields ($\sim$ 7 deg$^2$ each): the SXDS$+$XMM-LSS field, the COSMOS field, the ELAIS-N1 field, and the DEEP2-3 field.

\subsection{Excluding galaxies with inadequate surface brightness depth}\label{sec:selection_sblimit}

Due to spatial inhomogeneities in instrumental scattering\footnote{This effect has been largely corrected in the HSC-SSP data releases since DR2.}, large-scale flat-fielding uncertainties, and complex light contamination from nearby sources or scattering by cirrus dust in the MW, the local background noise level can vary from galaxy to galaxy. To maintain the quality of our stellar halo analysis, we exclude galaxies with insufficient surface brightness depth, defined as twice the local background rms noise. Below, we describe the procedure used to estimate the local background and its rms noise in the HSC-SSP Deep survey images of each galaxy.

Surface photometry in the low-surface-brightness regime is dominated not by the familiar pixel-to-pixel high-frequency noise (e.g., Poisson noise and pixel-scale flat-field errors), but by large-scale background inhomogeneities (i.e., low-frequency noise). To obtain a robust estimate of the local background and its noise, we follow the methodology of \citet{Gil_2005}. In particular, the low-frequency noise term is quantified by dividing the local background region around each galaxy into tens of approximately-equal-area sectors for statistical analysis. Further details are provided in Appendix \ref{sec:backnoise}. Based on this evaluation, we exclude 24 galaxies with $g$- and $r$-band surface brightness limits shallower than $29$ \magsqarc\ from subsequent analysis. The retained sample of 225 inclined central galaxies has a median $r$-band surface brightness limit of $\sim 29.9$ \magsqarc.

\subsection{The final sample of high-inclination, isolated central galaxies} \label{sec:highincl}

\begin{figure}[!ht]
  \centering
  \includegraphics[width=0.47\textwidth]{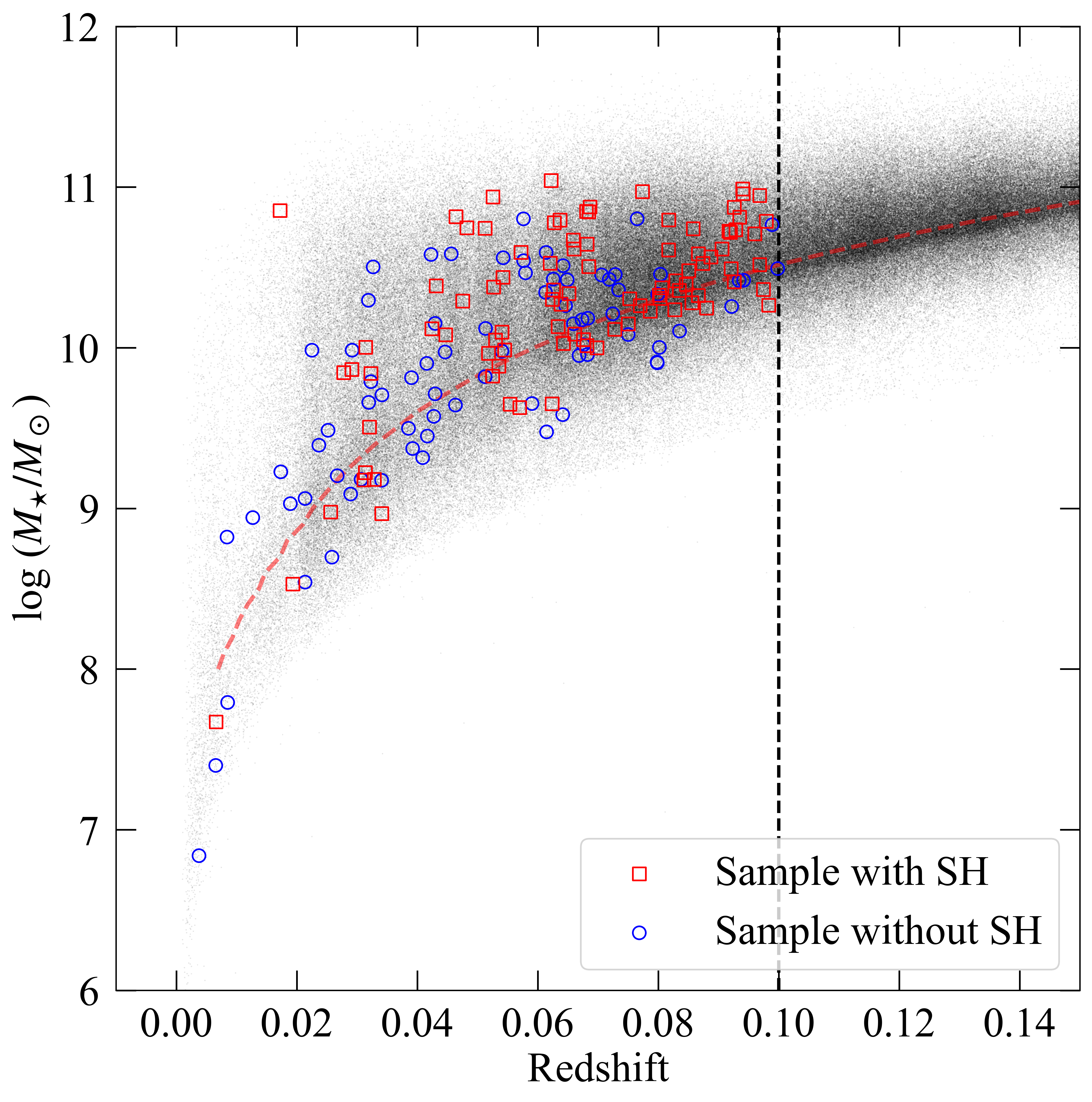} 
  \caption{Distribution of our galaxies on the stellar mass--redshift plane. The background gray dots show the parent spectroscopic sample from SDSS DR7 and GAMA DR4, limited to galaxies with $r<17.77$ mag. Blue open circles represent high-inclination central galaxies without stellar halo detections, while red open squares mark galaxies with stellar halo detections. The vertical dashed line indicates the upper redshift limit applied in our sample selection. The red dashed curve represents the 95\% completeness limit of our parent sample of central galaxies. Stellar masses plotted here are all taken from the parent catalogs. 
  }
  \label{Fig:Z-Mass}
\end{figure}

Finally, we perform isophotal fitting on galaxy images corrected for PSF-scattered light (Section \ref{sec:deconv}) to measure the average axial ratio ($q_{\rm disk}$) of isophotes within disk-dominated regions, defined as the radial interval near the minimum of the radial axial-ratio profile. From the 225 galaxies selected above, we then require $q_{\rm disk} < 0.4$, yielding a final sample of 169 high-inclination galaxies for analysis in this work. The high-inclination criterion of $q \lesssim 0.4$ has been adopted in previous studies \citep[e.g.,][]{Kautsch2006, Favaro2025}. We note that our conclusion in this work, such as the halo detection rate, does change if adopting a more restrictive threshold of $q_{\rm disk} < 0.3$.

\begin{figure*}[!ht]
  \begin{center}
    \includegraphics[width=1.0\textwidth]{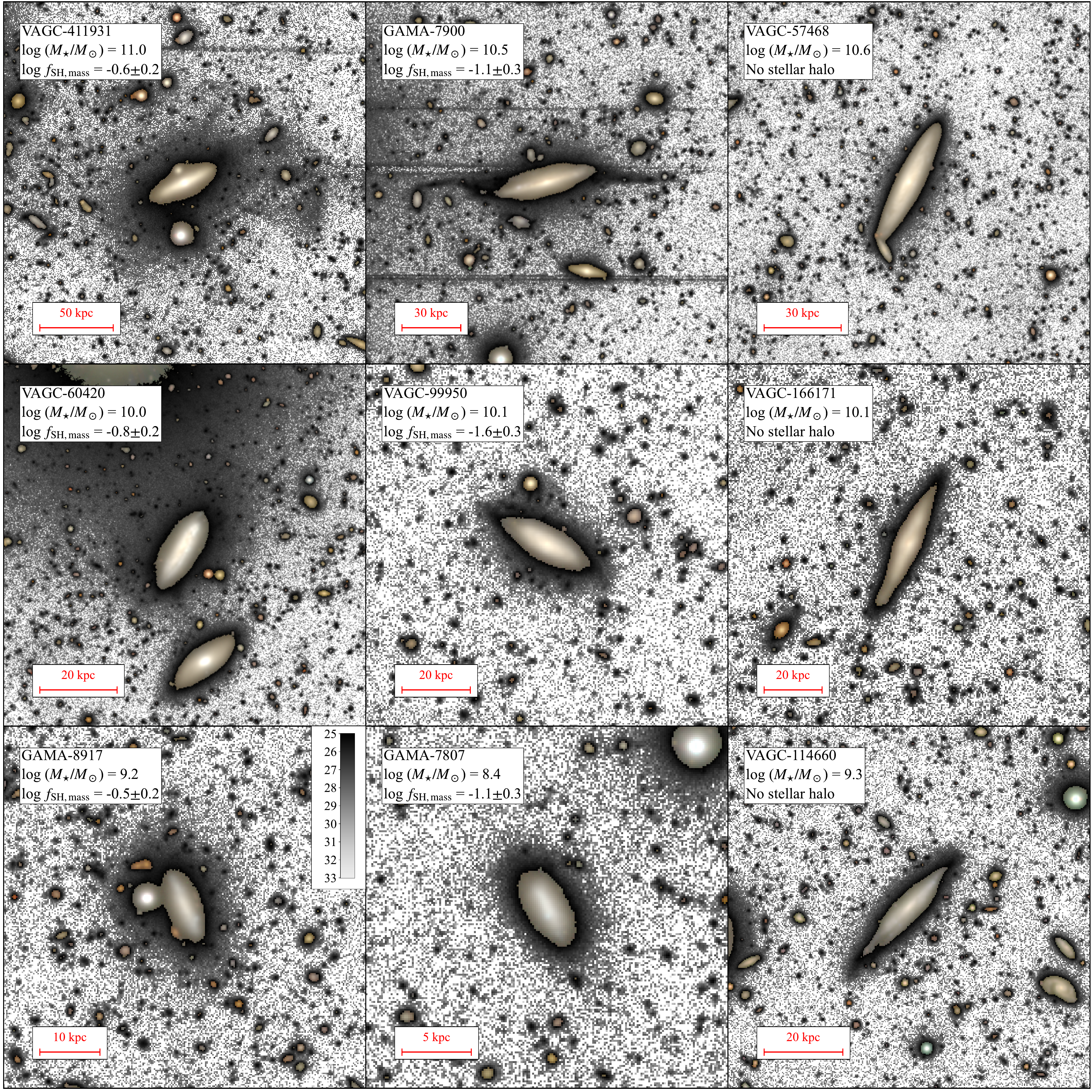}
    \caption{HSC-SSP Deep/UltraDeep images of nine representative galaxies in our final sample of high-inclination central galaxies. For each galaxy, the $g$, $r$, and $i$-band tricolor image of the high surface brightness central region is combined with the white-light grayscale image to reveal the lower surface brightness outskirts. Images are re-binned to enhance visibility of the low-surface brightness outskirts. The first row shows galaxies with MW-like stellar masses, $\log(M_{\star}/M_{\odot}) \sim (10.5,\,11.0)$. The second row displays galaxies with $\log(M_{\star}/M_{\odot}) \sim (10.0,\,10.5)$. The last row represents galaxies with the lowest stellar masses, $\log(M_{\star}/M_{\odot}) < 10.0$. The first column depicts galaxies with high stellar halo fractions, $\log f_{\rm halo} > -1$. The middle column shows galaxies with moderate stellar halo fractions, $\log f_{\rm halo} \sim (-2,\,-1)$. The last column shows galaxies without stellar halo detections. The color bar of the grayscale image in the lower-left panel is in units of \magsqarc.}
    \label{Fig:SampleSnap1}
  \end{center}
\end{figure*}

Figure \ref{Fig:Z-Mass} shows the redshift–stellar mass distribution of our high-inclination galaxy sample. The sample spans the full redshift and stellar mass range of the complete spectroscopic sample from SDSS and GAMA with $r$-band apparent magnitudes brighter than 17.77 mag. The 95\% completeness limit of stellar mass  as a functon of redshift for the parent spectroscopic sample is also indicated (as red dashed curve) in Figure \ref{Fig:Z-Mass}. Figure \ref{Fig:SampleSnap1} presents HSC-SSP cutout images of nine representative galaxies from our sample, selected to cover a range of stellar masses and stellar halo fractions. These nine galaxies will be used in the following sections to illustrate the image processing procedures and surface brightness profile extraction applied to the full sample.

\section{Methods} \label{sec:method}

In this section, we elaborate on the methods used to process HSC-SSP images and to extract the surface brightness profiles of the stellar halos of our galaxies.

\subsection{Image processing} \label{sec:imageprocess}

\subsubsection{Wide PSF construction and bright star subtraction} \label{sec:widepsf}

\begin{figure}[!ht]
  \centering
  \includegraphics[width=0.47\textwidth]{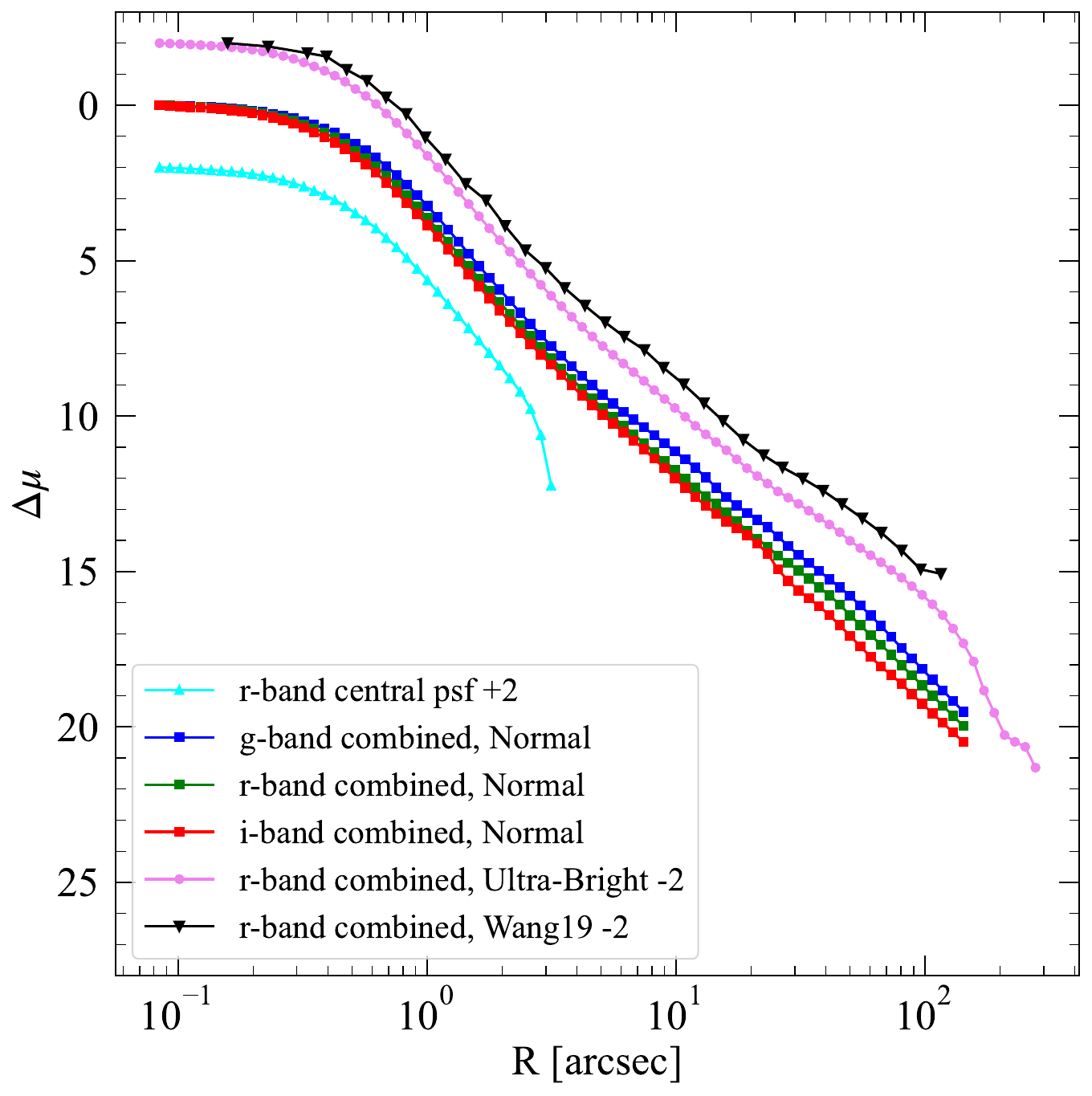} 
  \caption{Radial profiles of PSFs constructed for the HSC-SSP Deep/UltraDeep images. Cyan triangles show the $r$-band PSF profile provided by HSC-SSP PDR3, which captures only the PSF core due to atmospheric turbulence but misses the extended PSF wings. The combined ``normal'' wide-PSF profiles constructed in this work are shown as red squares for the $i$ band, green squares for the $r$ band, and blue squares for the $g$ band. The combined ``Ultra-Bright'' $r$-band wide-PSF profile is shown as violet circles. For comparison, black triangles represent the $r$-band wide-PSF from \citet{Wang_2019}. See Section \ref{sec:widepsf} for details.}
  \label{Fig:WidePSF}
\end{figure}

As demonstrated in previous studies \citep{deJong_2008, Tal_2011, Sandin_2014, Trujillo_2016, Wang_2019, Gilhuly_2022}, the PSF generally has extended wings that scatter light from bright sources (e.g., stars, galaxy bulges) over large angular scales, which can significantly bias surface photometry of extended diffuse structures such as galaxy stellar halos (see Figure \ref{Fig:SampleSnap1}). To obtain reliable surface brightness profiles in the galaxy outskirts, this scattered light contamination from the PSF wings must be carefully modeled and subtracted. The HSC-SSP survey team provides PSF images across the surveyed field, but these include only the central few arcseconds (see the cyan profile in Figure \ref{Fig:WidePSF}), dominated by atmospheric turbulence, while missing the extended wings of the PSF. Below, we describe our strategy for constructing wide-PSF models that capture the wings for the HSC-SSP Deep layer in the $g$, $r$, and $i$ bands. 

The PSF images provided by the HSC-SSP team are constructed from unsaturated stars, which do not have sufficient signal-to-noise to reach the extended PSF wings. We construct the wide PSF using a non-parametric method that combines stacked profiles of both unsaturated and centrally saturated bright stars, broadly following the procedure outlined in \cite{Wang_2019}.

\begin{figure*}[!ht]
  \begin{center}
    \includegraphics[width=1.0\textwidth]{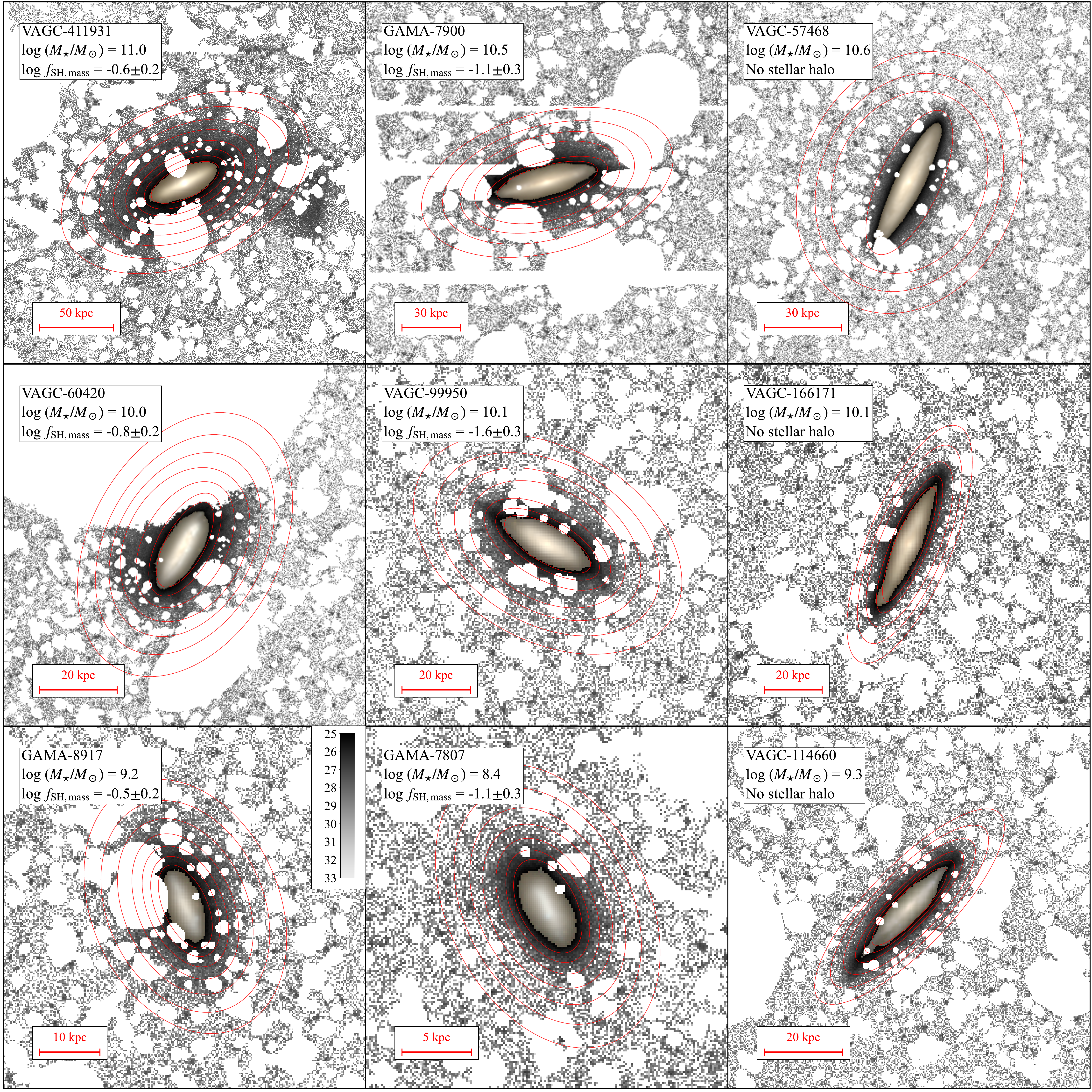}
    \caption{Similar to Figure \ref{Fig:SampleSnap1}, but after subtracting bright stars and masking contamination sources. Red ellipses show the isophotal contours in the low surface brightness outskirts. See Sections \ref{sec:widepsf} and \ref{sec:mask} for details.}
    \label{Fig:SampleSnap2}
  \end{center}
\end{figure*}

We use the Gaia EDR3 catalog \citep{Collaboration_2016, Collaboration_2021} to select bright stars for creating wide PSFs within the HSC-SSP Deep layer footprint. For all Gaia sources with $G < 20$ mag, we retrieve coordinates, parallaxes, proper motions, $G$ magnitudes, and BP$-$RP colors. To exclude potential contamination from compact background galaxies or AGNs, we discard sources with parallax $< 0.1$ mas (corresponding to $d > 10$ kpc), thereby constructing a relatively clean star sample for stacking. The selected stars are cross-matched with the HSC-SSP source catalog after correcting their coordinates for proper motion. Because the central pixels of the brightest stars are saturated in HSC images, the integrated fluxes reported by HSC-SSP cannot be used to normalize their images for stacking analysis. To overcome this, we calibrate empirical relations that convert Gaia $G$ magnitude to HSC $g$, $r$, and $i$-band PSF magnitude as a function of BP$-$RP color, using stars with $19 < G < 20$ mag that remain unsaturated in the HSC-SSP images.

We extract HSC-SSP Deep image cutouts centered on the above selected stars in the magnitude range of $5 < G < 18$ mag. To minimize contamination from background galaxies and other stars, we employ \textsc{SExtractor} to detect and mask the contaminants. We also subtract a sky background separately for each star image.

Brighter stars extend further into the PSF wings with higher signal-to-noise but may contain more saturated pixels near the center. Following the approach of previous studies \citep[e.g.,][]{Infante-Sainz2020}, we construct different radial ranges of the wide PSF using stars of different brightness and then combine them into a single wide PSF image. Specifically, we divide the sample of bright stars into four magnitude bins ($5 < G < 12$ mag, $12 < G < 14$ mag, $14 < G < 16$ mag, and $16 < G < 18$ mag), and stack normalized, contamination-masked cutouts of stars in each bin separately. Stacking is performed with a 5-sigma clipped median, and the wing regions are further smoothed with Gaussian kernels to enhance the signal-to-noise ratio. Finally, the PSF core provided by the HSC-SSP data team is merged with the four stacked images tracing PSF wings at different radial intervals, yielding the full wide PSF models in the HSC $g$, $r$, and $i$ bands.

The above four stacked PSF wing images generally align with each other at the overlap, non-saturated part, which makes the combination straightforward. However, we find that the wide-PSF models constructed from stars with $G > 12$ mag differ in their outermost regions from those built with stars with $G < 12$ mag (Figure \ref{Fig:WidePSF}). We therefore designate the former as the normal wide-PSF model and the latter as the Ultra-Bright wide-PSF model. The normal wide-PSF model is not sufficiently extended to reach scales beyond $R_{\rm p} \sim 80$ arcsec. To address this, we extrapolate the normal model out to $R_{\rm p} \sim 150$ arcsec by fitting a power law, $I_p \propto r^{-n}$, to its surface brightness profile in the range $50 < R_{\rm p} < 80$ arcsec \citep{Sandin_2014, Liu_2022}. Since all galaxies in our sample are fainter than $G = 12$ mag, we adopt the normal wide-PSF model to correct for PSF scattering from our target galaxies (see below). As shown in Figure \ref{Fig:WidePSF}, the resulting wide-PSF profiles are broadly consistent with those reported by \cite{Wang_2019} based on the HSC-SSP Wide survey images.

We subtract bright stars contaminating the galaxy images by using the wide-PSF models and Gaia catalog, down to a surface brightness level of $\mu \sim 30$ \magsqarc. Specifically, the Ultra-Bright wide-PSF models are applied to stars with $G < 10$ mag, the normal wide-PSF models to stars with $G > 12$ mag, and stars with $10 < G < 12$ mag are modeled using the average of the normal and Ultra-Bright wide-PSF models. Since subtraction residuals often remain in the central regions of bright stars, we mask all areas brighter than $\mu < 27.5$ \magsqarc and exclude them from further analysis. Figure \ref{Fig:SampleSnap2} shows cutouts of the nine representative galaxies after bright-star subtraction, illustrating the effectiveness of this step in our image processing.

\subsubsection{Errors introduced by spatial variations of PSF} \label{sec:psferror}
While it is necessary to stack a large number of stars to obtain high-quality PSF models, applying a stacked PSF model to the whole field may carry an uncertainty incurred by potential spatial variation of the PSF. For approximately 140 galaxies in our sample, we identified at least five nearby bright stars (with $12 < G < 14$ mag) within a 10-arcmin radius, enabling the construction of a moderate-quality average PSF model extending to 30" for each galaxy. Analysis of these 140 average PSFs reveals that the rms scatter of the normalized profiles increases from 11\% at 20" to 15\% at 30".

For a galaxy at a redshift 0.08 (typical for our MW-mass galaxies), a physical scale of 40 kpc (the outermost galactocentric radius adopted for stellar halo measurement, see the next sections) corresponds to $\sim$ 26". Within this angular radius, the rms scatter of the PSF is $\leq$ 13\%. As demonstrated by \cite{Wang_2019} using HSC-SSP wide-field images, PSF-scattering contributes less than 50\% (10\%) of the observed light at a radius of 40 kpc (10 kpc) for galaxies with $\log(M_{\star}/M_{\odot}) \geq 9.2$. Consequently, subtracting the PSF-scattered light introduces an uncertainty of at most 1\% (10\% $\times$ 13\%) and 6.5\% (50\% $\times$ 13\%) to the flux measurement at 10 kpc and 40 kpc, respectively. These small uncertainties are negligible compared to other sources of error.

\subsubsection{Contaminant source masking for target galaxies} \label{sec:mask}

To accurately extract the light distribution of stellar halos in our target galaxies, it is necessary to first mask foreground and background contaminating objects, in addition to the bright-star subtraction and masking described above. This step is nontrivial because such contaminants span a wide range of angular sizes, morphologies, and separations from the target galaxies. We use \textsc{SExtractor} to detect contaminants in the HSC $g$, $r$, and $i$-band cutout images centered on each galaxy, adopting several parameter combinations listed in Table \ref{tab:SEparms}. The three key \textsc{SExtractor} parameters are the source detection threshold (\texttt{DETECT\_THRESH}), the minimum number of contiguous pixels above the threshold (\texttt{DETECT\_MINAREA}), and the mesh size for background estimation (\texttt{BACK\_SIZE}). Segmentation maps generated with these settings are then used to mask all objects except the central galaxy itself. Below we briefly outline the rationale for our mask generation strategy.

In general, a low \texttt{DETECT\_THRESH} with a large \texttt{BACK\_SIZE} is required to capture the full extent of an object, but this increases the risk of spurious detections from background noise. To mitigate this, we pair the lowest \texttt{DETECT\_THRESH} with a relatively large \texttt{DETECT\_MINAREA} (\texttt{segmap\_05} in Table \ref{tab:SEparms}) and further require that the same objects also be detected with a higher \texttt{DETECT\_THRESH} and smaller \texttt{DETECT\_MINAREA} (\texttt{segmap\_15}). However, this combination does not identify all contaminants, particularly those projected close to the target galaxies. To recover these, we additionally use a series of higher \texttt{DETECT\_THRESH} values and smaller \texttt{BACK\_SIZE} (\texttt{segmap\_20}, \texttt{segmap\_30}, and \texttt{segmap\_40}) to detect objects missed in \texttt{segmap\_05} and \texttt{segmap\_15}.

\begin{deluxetable}{lccc}
\tablecaption{\textsc{SExtractor} parameters for masking\label{tab:SEparms}}
\tabletypesize{\footnotesize}  
\tablewidth{0pt}               
\tablehead{%
  \colhead{Name} &
  \colhead{\texttt{DETECT\_THRESH}} &
  \colhead{\texttt{DETECT\_MINAREA}} &
  \colhead{\texttt{BACK\_SIZE}}}
\startdata
\texttt{segmap\_05} & 0.5 & 25 & $2 \times R_{27}$ \\
\texttt{segmap\_15} & 1.5 & 4  & $2 \times R_{27}$ \\
\texttt{segmap\_20} & 2.0 & 4  & $R_{27}$ \\
\texttt{segmap\_30} & 3.0 & 4  & $R_{27}$ \\
\texttt{segmap\_40} & 4.0 & 4  & $0.5 \times R_{27}$ \\
\enddata
\tablecomments{The table lists the most relevant \texttt{SExtractor} parameters used to create contaminant source masks. In the last column, $R_{27}$ denotes the major-axis radius of the isophote with $\mu_{r}\,=\,27$\,\magsqarc. See Section \ref{sec:mask} for details.}
\end{deluxetable}

\begin{figure}[!ht]
  \centering
  \includegraphics[width=0.47\textwidth]{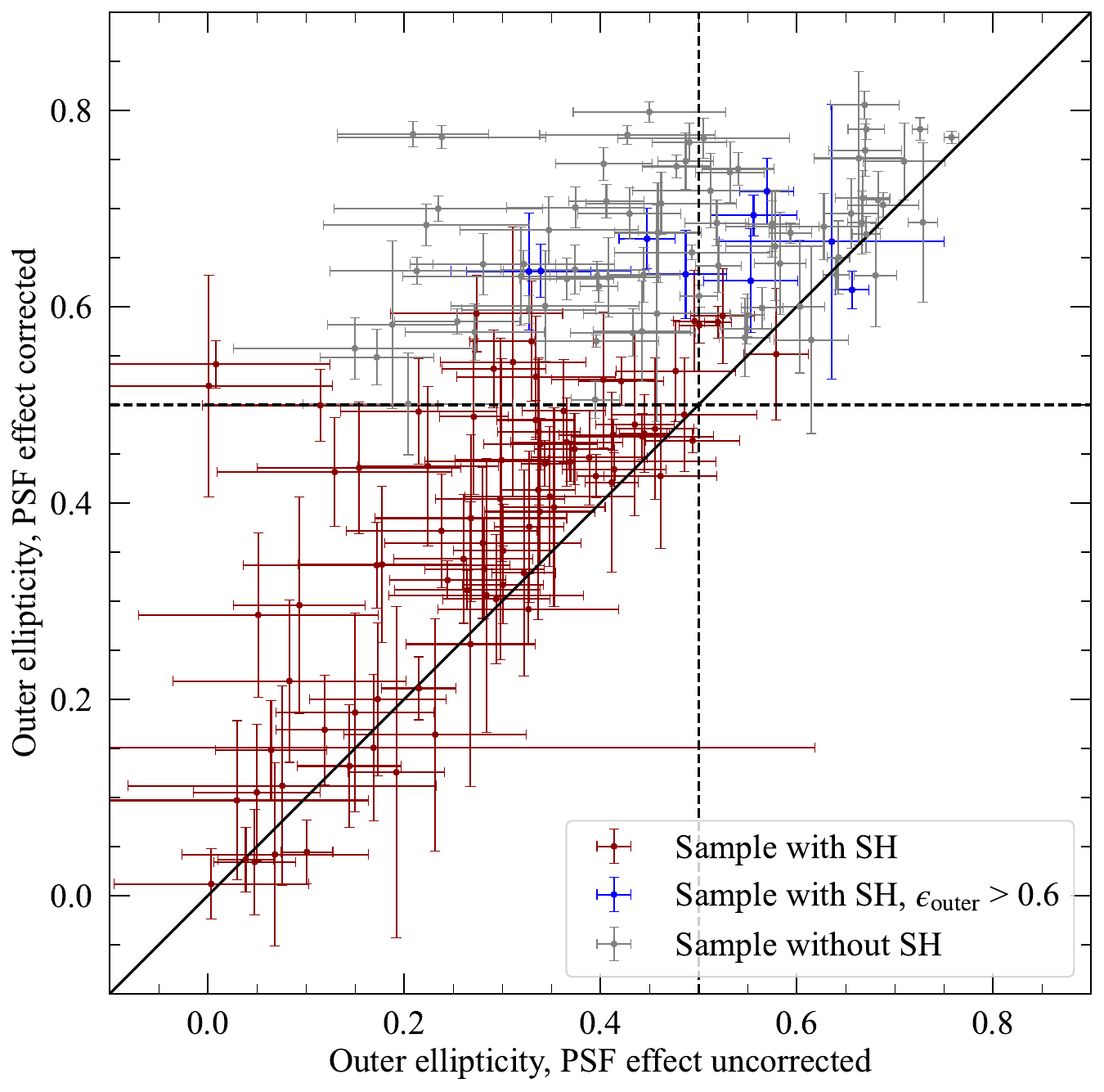} 
  \caption{Comparison of the outskirts ellipticities measured on images with and without subtracting the PSF-induced scattered light. Dark-red circles represent galaxies satisfying the ellipticity threshold (after correcting for the PSF effect) for stellar halo identification, while blue circles represent galaxies with power-law surface brightness profiles in the outskirts but do not satisfy the ellipticity requirement for halo identification (see Section \ref{sec:indenthalo} for details). Gray circles indicate galaxies without stellar halo detection. This comparison demonstrates the necessity of subtracting PSF-induced scattered light for reliable ellipticity measurements and stellar halo selection.} \label{Fig:SubsampleSelect}
\end{figure}

The segmentation maps returned by \textsc{SExtractor} cannot capture the full extent of detected objects, even at the lowest detection threshold. To compensate, we expand the segmentation boundaries by 3 and 2 pixels for objects detected in the first two and last three \textsc{SExtractor} runs listed in Table \ref{tab:SEparms}, respectively. For larger contaminating companions (HSC $r < 20$ or detected areas comparable to the central galaxy), we further extend the elliptical apertures from \texttt{segmap\_15} to 2.5 times the R90 (the radius enclosing 90\% of the flux). All segmentation maps are then visually inspected and adjusted (if necessary) to ensure that contaminants are fully masked, while spiral arms or other substructures of the target galaxies (if any) are not mistakenly masked, with additional manual masking applied when necessary. The final masks used in our analysis combine these segmentation maps with the bright star masks described in Section \ref{sec:widepsf}.

\subsubsection{Correction of galaxy light profiles for PSF scattering} \label{sec:deconv}

\begin{figure*}[!ht]
  \begin{center}
    \includegraphics[width=1.0\textwidth]{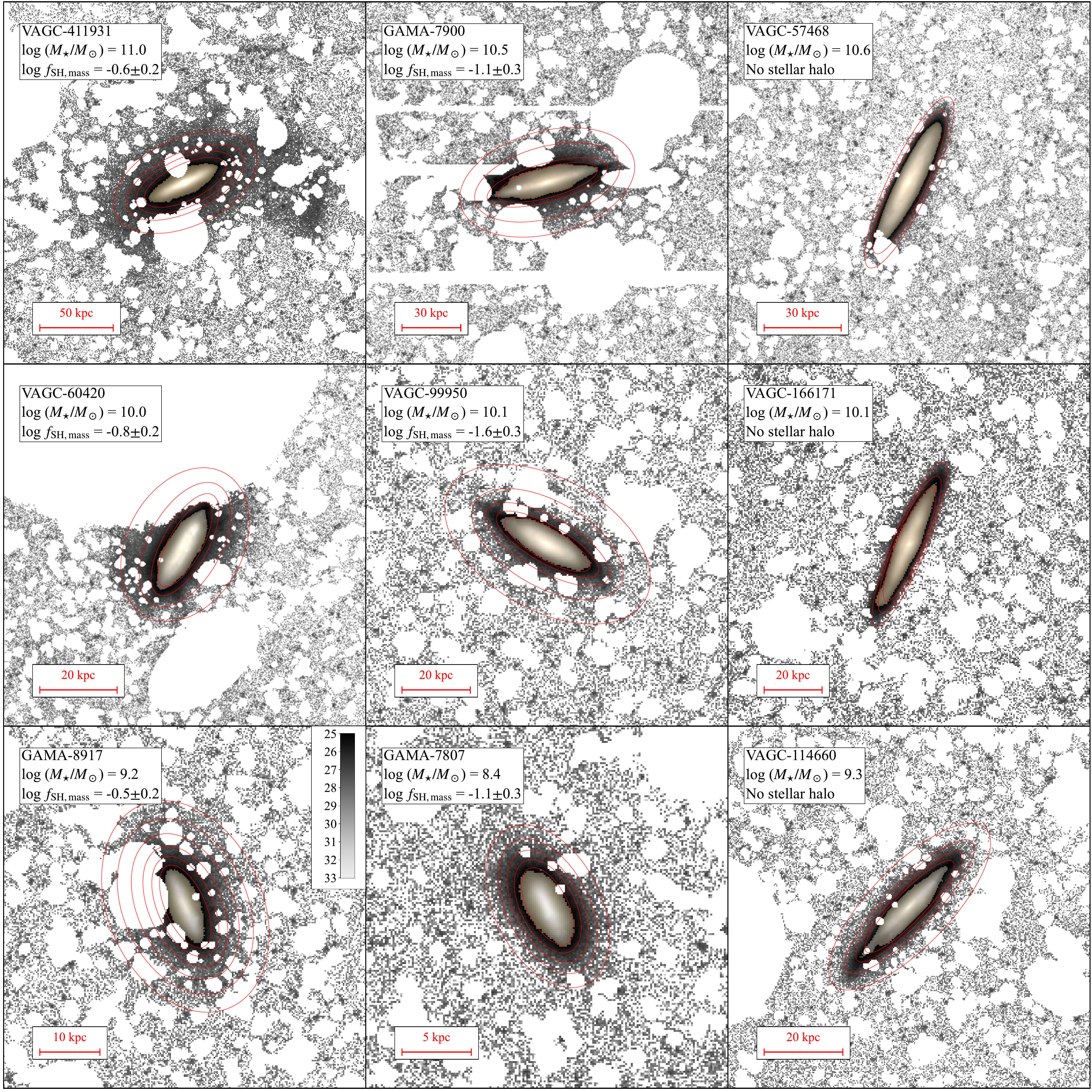}
    \caption{Similar to Fiugre \ref{Fig:SampleSnap1}, but after subtracting bright stars, masking out contaminating sources and subtracting PSF-induced scattered light. Red ellipses depict the isophotal contours in lower surface brightness outskirts.}
    \label{Fig:SampleSnap3}
  \end{center}
\end{figure*}

The extended wings of the PSF can scatter light from the high surface-brightness central regions of galaxies into their stellar halos. If uncorrected, this effect can lead to an overestimation of both the stellar halo fraction ($f_{\rm halo}$; \citealt{Wang_2019}) and the halo axial ratio ($q$). To accurately measure the stellar halos of our target galaxies, we adopt a method similar to that of \cite{Szomoru_2010} and \cite{Gilhuly_2022} to remove light scattered from the galaxy centers. Specifically, we fit PSF-convolved galaxy models to the observed images with \texttt{GALFIT} \citep{Peng_2002, Peng_2010}, and derive the scattered light as the difference between the best-fit models with and without PSF convolution. The most relevant details of this procedure are described below.

In general, we model galaxy disks using either an exponential profile (a Sersic model with $n=1$) or an edge-on disk model. If a prominent bulge is present, an additional Sersic component is included. Because stellar halos contribute only a negligible fraction of the light relative to the disk and bulge in the bright stellar main body, particularly along the major axis, such one- or two-component modeling with \texttt{GALFIT} is not affected by the presence of stellar halos. Upon inspecting the brightness profiles, we find that about 25\% of the HSC-SSP Deep $i$-band images of our target galaxies suffer from saturation in their centers, with a few cases also showing saturation in the $r$-band. Since the single-exposure time of the Wide survey is much shorter than that of the Deep survey, we use the corresponding Wide survey images for model fitting in these cases. We construct appropriate wide-PSF models applicable to the Wide survey images by retrieving the PSF core images provided by the HSC-SSP Wide survey and combining them with the normal wide-PSF model, joined smoothly at radii of $\sim$1–1.5 arcsec.

\texttt{GALFIT} does not provide a fully holistic framework for structural analysis, and simultaneously fitting multi-component models can sometimes yield non-physical solutions, as the model selection is primarily by minimizing fitting residuals. To obtain reliable disk and bulge models across all bands, we implement a three-step process. In the first step, we mask the potential bulge region and fitted the disk model to the disk region. In the second step, we fix the disk model and used a Sersic model to fit the bulge. The Sersic index for the bulge model is set between 2 and 4. Considering that the $i$-band image of HSC-SSP Wide survey has the best seeing quality ($\sim 0.6$ arcsec), we use the $i$-band images to determine the morphological parameters of the bulge. In the third step, we fix all the morphological parameters of the bulge model (center, $R_{e, \rm bulge}$, $n$ index, $q$, and PA) and the disk model's center and position angle, leaving other parameters free for fitting the $g$, $r$, and $i$-band images, obtaining suitable disk and bulge models. Although accurate measurements of bulge sizes ($R_{e, \rm bulge}$) are challenging for the most distant galaxies in our sample, the correction for PSF effects remains robust, as demonstrated by \citet{Szomoru_2010} and \citet{Szomoru_2012}, provided that the residuals do not exhibit systematic biases.

After completing the model fitting, we visually inspect the residuals to identify cases showing systematic deviations from the observation in the central regions. We refine these models by adjusting the three-step procedure mentioned above. For example, in some galaxies with a central deficit instead of excess, the exponential disk fitting caused systematic negative residuals in the central region, so we use a Sersic model with an index of slightly smaller than 1 to fit the entire galaxy main body. In cases where the bulge occupies a significant fraction of the galaxy, the default bulge mask is insufficient, leading to significant systematic deviations of the model from the observed profile in the center. To improve the fitting to these galaxies, we free both the bulge and disk model parameters in the second step to achieve an optimal fitting result. For few galaxies suffering from central saturation in the HSC-SSP Wide field image in the $i$-band, we fit the bulge parameters using the $r$-band image and then used the empirical bulge color of unsaturated galaxies ($r - i \sim 0.41$) to convert the $r$-band bulge model to the $i$-band. We mask the saturated central regions of the $i$-band image and fix the bulge component while fitting the disk component.

Using the best-fit two-dimensional galaxy model parameters (disk and bulge components, if present) of our target galaxies, we correct for PSF effects in the Deep field images. With \texttt{GALFIT}, we generate galaxy model images both with and without convolution by the normal wide-PSF model; their difference represents the light scattered by the PSF. Subtracting this difference image from the Deep field data yield the PSF-corrected galaxy images. For galaxies with saturated central regions in the Deep field, the saturated pixels are replaced with the unconvolved galaxy model.

Figure \ref{Fig:SampleSnap3} shows processed images of the same nine representative galaxies shown in Figures \ref{Fig:SampleSnap1} and \ref{Fig:SampleSnap2}, after bright star removal, contaminant masking, and PSF-scattered light subtraction. The change in the light distribution of galaxy outskirts, compared to Figures \ref{Fig:SampleSnap1} and \ref{Fig:SampleSnap2}, is noticeable. A comparison of the outskirts ellipticities measured on images with and without subtraction of PSF-induced scattered light is shown in Figure \ref{Fig:SubsampleSelect}, highlighting the importance of correcting for PSF effects when measuring stellar halos.

\subsection{Extracting galaxy surface brightness profiles}\label{sec:surfprofile}

\begin{figure*}[!ht]
  \begin{center}
    \includegraphics[width=1.0\textwidth]{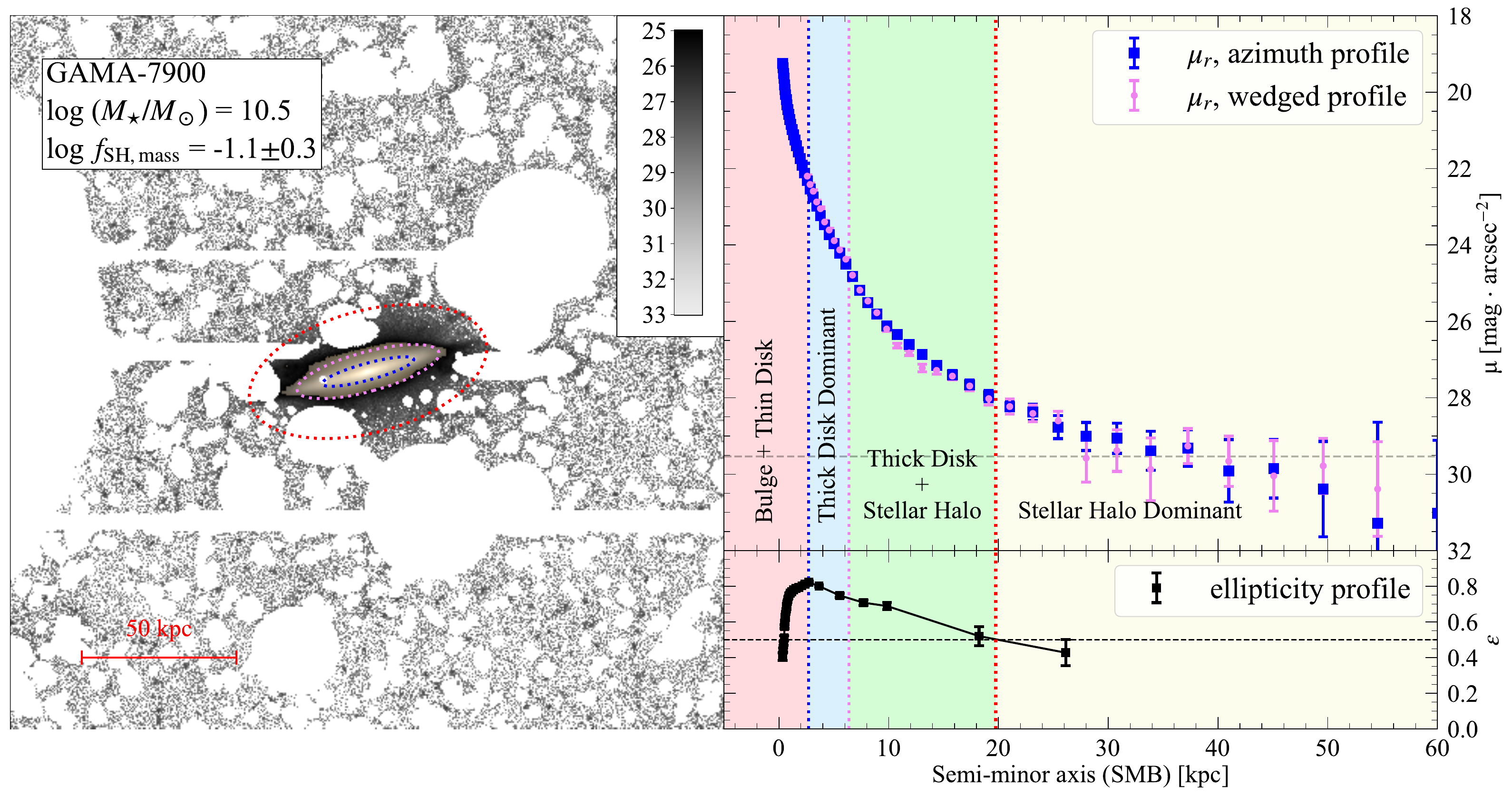}
    \caption{An example of identifying the stellar halo-dominated regions. \textbf{Left:} A composite image of a galaxy, produced in a similar way to that shown in Figure \ref{Fig:SampleSnap3}. Colored dotted ellipses mark the division of the radial ranges dominated by different galaxy components (i.e., bulge, thin disk, thick disk, and stellar halo). \textbf{Upper right:} Radial surface brightness profiles along the minor axis. Blue squares with error bars show the azimuthally median profiles, while violet circles represent the median profiles measured within $60^\circ$ wedges around the minor axis above and below the disk plane. The horizontal dashed line indicates the $2\sigma$ surface brightness limit. Vertical dotted lines mark the same division radii as represented by the ellipses in the left panel, using the same color scheme as the ellipses. \textbf{Lower right:} Isophotal ellipticity profile. All radial profiles are plotted as a function of galactocentric distance along the semi-minor axis (SMB) in kpc. See Section \ref{sec:goldensample} for details.}
    \label{Fig:Detection}
  \end{center}
\end{figure*}

We use the \texttt{photutils.isophote.ellipse} task to fit elliptical isophotes and extract radial profiles of the azimuthal median surface brightness. The isophotal fitting is performed iteratively. First, we fit elliptical isophotes to white-light images, created as a weighted average of the $g$, $r$, and $i$ bands without PSF correction, allowing all geometric parameters to vary with radius. This step determine the average galaxy center based on the best-fit isophotes at intermediate disk radii.

Next, using white-light images produced from PSF-corrected multi-band images, we perform ellipse fitting with the center fixed to the value obtained previously, while allowing the axis ratio ($q$) and position angle (PA) to vary with radius. We ran \texttt{ellipse} with \texttt{integrmode}=median and a fractional \texttt{step}=0.2, and visually inspected the best-fit isophotes to ensure convergence and a smooth radial variation of $q$ and PA. In the third round of \texttt{ellipse} fitting, in addition to fixing the galaxy center, we also fix the PA to the average value measured at the intermediate disk radii where the PA remains nearly constant. Occasional abrupt jumps in $q$ are manually corrected by replacing them with linearly interpolated values from adjacent isophotes. The convergence threshold and maximum gradient error parameters in \texttt{ellipse} are kept at their default settings. We obtain high-quality isophotal fits for the relatively bright disk region in the third \texttt{ellipse} run. In the fourth \texttt{ellipse} run, we increase the fractional \texttt{step} parameter to 0.3 in order to derive reliable ellipticities for the extended and faint outskirts, where the third run fails to converge. The isophotal fits for these faint outer regions are then merged with the converged fits for the bright disk region obtained from the third run. In the outermost regions, where the signal-to-noise ratio is insufficient for convergence, the geometric parameters are fixed to those of the nearest converged isophotes. Lastly, to achieve finer radial sampling of the surface brightness profiles than that used for constraining the isophotal geometric parameters, we reduce the fractional step size by a factor of two, linearly interpolating the best-fit radial variation of the ellipse parameters obtained from the previous fitting.

Finally, we measure the isophotal radial brightness profiles from the PSF-corrected $g$, $r$, and $i$ band images separately, fixing all geometric parameters to the previously determined values. Using the \texttt{build\_ellipse\_model} task, these radial profiles are converted into two-dimensional galaxy models. Masked pixels in the galaxy images are replaced with the model values to create ``clean'' images. These clean images are then used to measure the accumulated flux within elliptical apertures along the semi-major axis (SMA), enabling the construction of the curve of growth (COG) and the derivation of quantities such as total flux, $R_{\rm e, maj}$, and concentration parameters in each band.

\begin{figure*}[!ht]
  \begin{center}
    \includegraphics[width=1.0\textwidth]{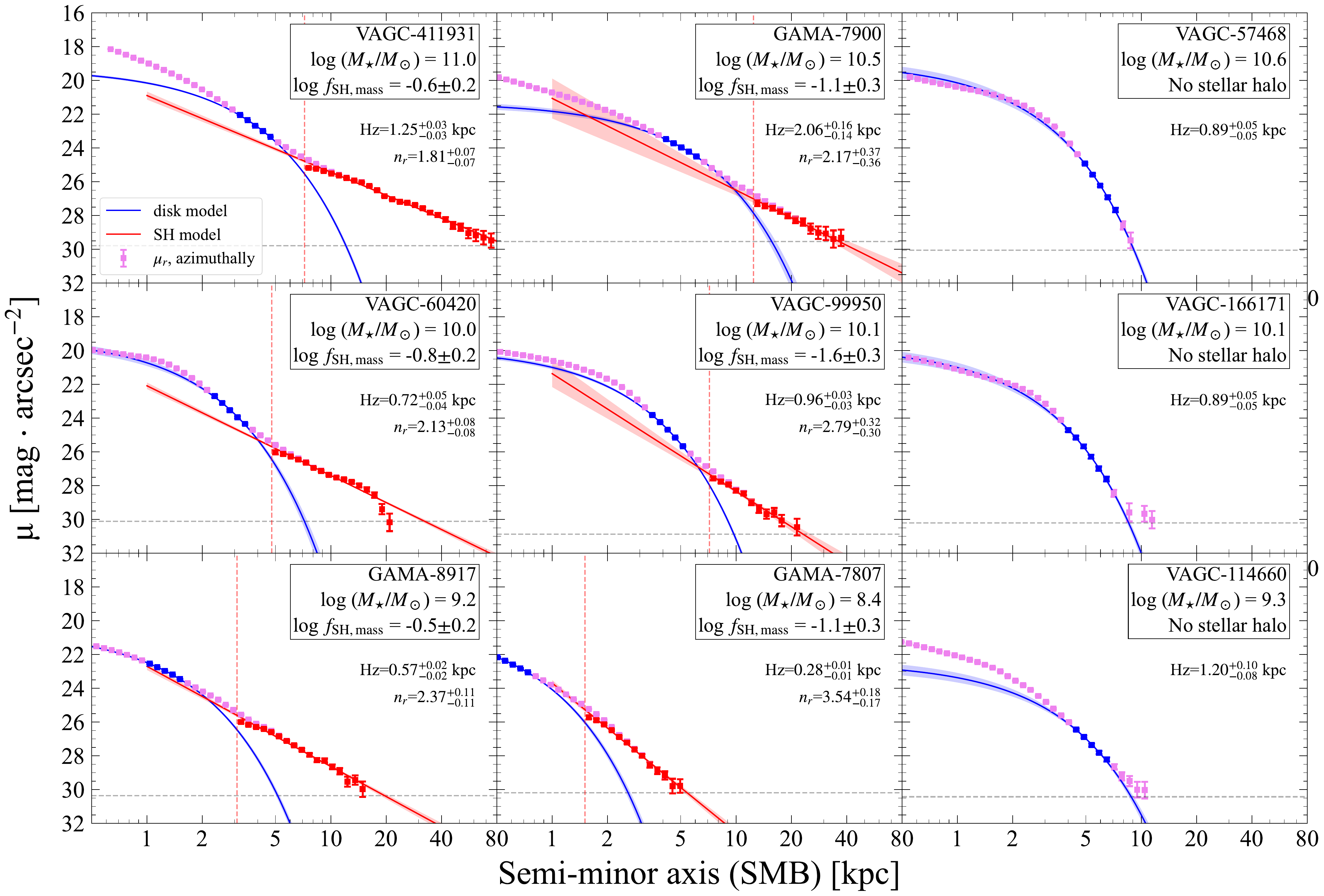}
    \caption{Examples of stellar halo measurements from surface brightness profiles. Each panel shows one of the same galaxies presented in Figure \ref{Fig:SampleSnap3}. Violet squares with error bars represent the $r$-band azimuthally median surface brightness profiles along the minor axis. Blue squares indicate the outer disk (thick disk)–dominated region used for fitting an exponential model, with the best-fit exponential overplotted as blue curves. Red squares show the disk-subtracted surface brightness profiles used for power-law fitting (red curves) of the halo-dominated region. The red vertical dashed line marks the inner boundary of the stellar halo–dominated region adopted for halo profile fitting, and the gray horizontal dashed line indicates the 2$\sigma$ surface brightness limit. The best-fit thick disk exponential scale height and the stellar halo power-law index are also indicated in each panel. See Section \ref{sec:fullsample} for details.}
    \label{Fig:Measurement}
  \end{center}
\end{figure*}

\begin{figure*}[!ht]
  \begin{center}
    \includegraphics[width=1.0\textwidth]{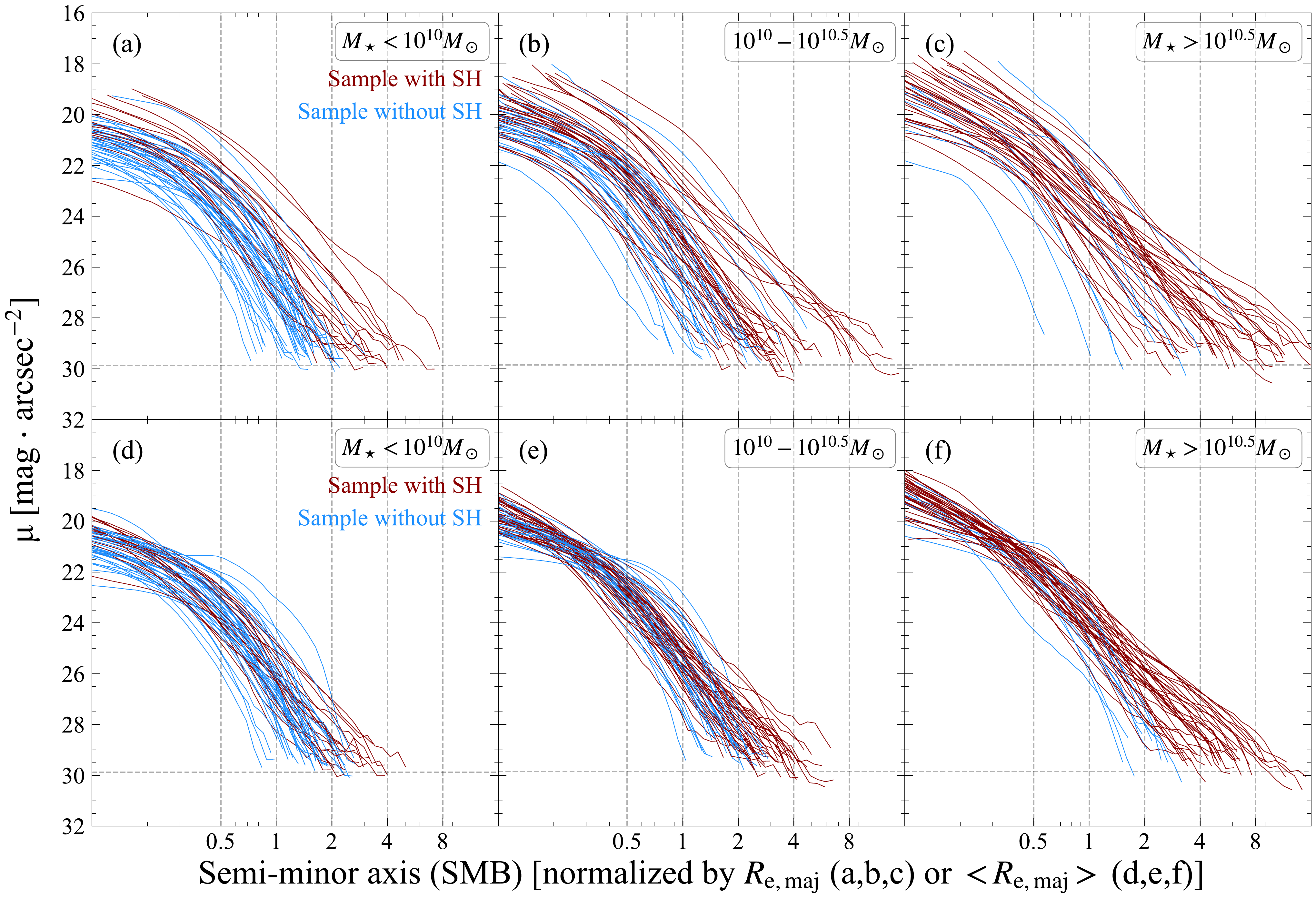}
    \caption{The $r$-band azimuthally median surface brightness profiles along the minor axis of all sample galaxies, shown in units of $R_{\rm e, maj}$ (upper panels) or $<R_{\rm e, maj}>$ (lower panels). Galaxies are grouped by stellar mass range in different columns, as indicated in each panel. Here, $R_{\rm e, maj}$ denotes the half-light radius along the major axis, and $<R_{\rm e, maj}>$ represents the mass-dependent average $R_{\rm e, maj}$ predicted by a piecewise log-log linear fit to the $R_{\rm e, maj}$–stellar mass relation of our galaxies. Dark red profiles indicate galaxies with stellar halo detections, while dodger blue profiles indicate galaxies without stellar halo detections. This figure demonstrates that surface brightness profiles normalized by the average major-axis half-light radius exhibit substantially smaller scatter than those normalized by each galaxy's own half-light radius. See Section \ref{sec:massestimate} for details.}
    \label{Fig:Smb_norm}
  \end{center}
\end{figure*}

\subsection{Stellar halo identification and measurement} \label{sec:indenthalo}

Stellar halos are distinguished from the rotation-dominated, flattened galaxy disks by their dispersion-dominated and nearly spheroidal morphology. In high-inclination galaxies, stellar halos can be straightforwardly identified in galaxy images through their large minor-to-major axis ratios, $q_{\rm intr}$ (or equivalently, small ellipticity $\epsilon_{\rm intr} = 1 - q_{\rm intr}$). The isophotal fitting of galaxy images (Section \ref{sec:surfprofile}) provides the radial variation of $q_{\rm intr}$, enabling the identification of radial ranges dominated by the stellar halo.

Previous studies of several Local Volume galaxies based on star counts found that nearly all stellar halos exhibit axis ratios $q$ in the range of $\sim 0.5-0.8$ (see references in the introduction), and $q$ generally increases with galactocentric radius. Using Illustris simulations, \cite{Elias2018} demonstrates that over 95\% of MW-mass galaxies have stellar halos with an intrinsic minor-to-major axis ratio $> 0.5$. This fraction would be even higher when accounting for projection effects. The typical axis ratios of stellar halos have little overlap with those of thick disks in spiral and S0 galaxies \citep[$q \lesssim 0.4$; e.g.,][]{Favaro2025}, thereby permitting a clear geometric separation. 

\subsubsection{The most prominent stellar halos in the sample }\label{sec:goldensample}

Given the distinct morphologies of stellar halos compared to galaxy disks, we begin our halo identification by focusing on the most prominent stellar halos in the sample. These serve as a guide for developing a more complete identification of halos in our sample. To select the most prominent halos, we impose two criteria: 1) the halo-dominated radial range satisfies $q > 0.5$ (or equivalently, $\epsilon < 0.5$), and 2) the outermost boundary either extends to more than twice the inner boundary or includes no less than seven data points with S/N $>$ 2. The latter criterion ensures a robust discrimination of functional forms of the surface brightness profiles. Together, these two criteria yield a sample of 49 galaxies.

An example illustrating the differentiation of the stellar halo from other galaxy components is shown in Fig.~\ref{Fig:Detection}. In addition to the azimuthally median surface brightness profiles, the figure also shows median profiles measured within $60^\circ$ wedges centered on the minor axis above and below the disk plane. The wedge profiles provide a cleaner view of the halo, as they are less affected by potential disk contamination near the major axis, albeit at the cost of lower signal-to-noise ratio. Since we do not find significant differences between the surface brightness profiles from the two methods, especially in disk or stellar halo dominant regions, we adopt the azimuthally median profiles for the following analysis.

As exemplified in Fig.~\ref{Fig:Detection}, the galaxy is divided into radial ranges dominated by different structural components. The innermost division line (or ellipse) corresponds to the radius of maximum ellipticity, marking the maximum relative contribution of the thin disk. Inside this radius, the bulge component (if present) gradually changes the isophotal ellipticity toward smaller radii, while beyond it the contribution from the thick disk progressively increases. The outermost division line (or ellipse) is set at $\epsilon = 0.5$, corresponding to the conservative inner boundary of stellar halo-dominated regime. The intermediate division line (or ellipse) marks the radius where the surface brightness profile begins to deviate from the exponential fit at smaller radii, signifying the regime where the stellar halo starts to make a noticeable contribution.
 
We model the surface brightness profiles of stellar halo–dominated regions as a function of minor-axis radius using three functional forms: a single power law (SPL), a broken power law (BPL), and an exponential profile. To avoid the profile fitting being dominated by the densely sampled inner regions, we rescale the measurement variances of surface brightness at each radius by the reciprocal of the number of valid pixels in the corresponding elliptical ring. The uncertainties of the model parameters are then estimated with the Markov Chain Monte Carlo (MCMC) method, implemented via the \texttt{emcee} sampler. Model selection is performed using the Bayesian Information Criterion (BIC), which balances goodness of fit against model complexity by penalizing the number of parameters. Out of the 49 galaxies in our sample, 38 are best described by the SPL model, 6 by the BPL model, and only 5 by the exponential model. The prevalence of power-law profiles is consistent with previous findings for the MW and other nearby galaxies, where more sophisticated halo-star selection techniques have been applied. This agreement not only validates the minor-to-major axis ratio as a robust parameter for identifying stellar halo–dominated regions, but also confirms that stellar halos generally follow power-law radial profiles.

To obtain more precise measurements of stellar halos, we model the surface brightness in the thick-disk–dominated radial ranges with an exponential profile as a function of minor-axis distance, and subtract the best-fit exponential from the observed profile. The thick-disk–dominated region is identified as the radial range exhibiting exponential behavior beyond the peak of the ellipticity radial profile. In cases where no clear exponential radial segment is found beyond the peak, the fitting is performed using data near the peak. The $r$-band surface brightness profiles for the nine representative galaxies introduced before, along with the corresponding disk-subtracted profiles, are shown in Figure~\ref{Fig:Measurement}. We find that the surface brightness profiles of the 49 most prominent stellar halos remain largely unaffected by the disk component at radial ranges with $q$ $\geq$ 0.4 (or $\epsilon$ $\leq$ 0.6), retaining their power-law form. 

\subsubsection{The full sample of stellar halos}\label{sec:fullsample}

Based on the lessons learned from the most prominent halos, we establish three joint detection criteria to identify stellar halo in the full sample: 1) the outer radial range satisfies $q$ $>$ 0.5 (or $\epsilon$ $<$ 0.5); or 2) the outer radial range satisfies $q$ $>$ 0.4 (or $\epsilon$ $<$ 0.6), and follows a power-law radial decline along the minor axis; or 3) the outskirts remain at $q$ $<$ 0.4, but exhibit a power-law radial decline along the minor axis. Applying these criteria, we identify stellar halos in a total of 93 galaxies. Among these, 69 galaxies meet the first criterion, 15 galaxies satisfy the second, and 9 galaxies fulfill the third. We note that all stellar halos identified using the first criterion can be well described by a power-law radial profile.

To consistently define an inner boundary of the halo-dominated region in the galaxies identified above, we adopt the minor-axis radius where the halo contribution starts exceeding more than half of the observed light, or equivalently where the surface brightness decreases by less than 0.75 \magsqarc after disk subtraction. In practice, we impose a stricter criterion that the surface brightness decrease be less than 0.5 \magsqarc, allowing for a 2$\sigma$ margin, where $\sigma$ corresponds to the typical measurement uncertainty of 0.12 \magsqarc near the halo boundaries of our galaxies. This definition of inner boundaries of stellar halo-dominated regime is adopted to estimate stellar halo properties in following sections. In Figure~\ref{Fig:Measurement}, the inner boundaries of the nine representative galaxies are each marked by a red vertical dashed line.

Given the overwhelming dominance of single power-law profiles among the 49 most prominent halos, and the limited surface brightness dynamical range in the remaining galaxies of the full sample, we adopt a single power-law model to fit the disk-subtracted stellar halo profiles of all 93 galaxies for consistency, using the same method described in Section~\ref{sec:goldensample}. The single power-law model is given by:
\begin{equation}
    I = I_0 \times \left(\frac{\rm{SMB}/{\rm{kpc}}}{\rm{10\,kpc}}\right)^{-n},
\end{equation}

where $n$ is the power-law index, $\rm{SMB}/{\rm{kpc}}$ is the minor-axis radius in kilo-parsecs, and $I_0$ is the flux at the arbitrarily chosen characteristic radius 10\,kpc. Among the 93 galaxies with stellar halo detection, 78 have at least five data points in the stellar halo region in $r$-band, allowing for a robust single power-law fit. The remaining 15 galaxies have too few data points to constrain both $I_0$ and $n$. For the 78 well-sampled halos, we therefore fit both $I_0$ and $n$, while for the 15 sparsely sampled cases, we adopt the posterior distribution of $n$ derived from the well-sampled halos as a prior during the fitting.

\subsubsection{Estimation of total stellar halo mass}\label{sec:massestimate}

Stellar halos are expected to extend continuously into the central regions of galaxies, but their profiles can only be reliably constrained in the outskirts where the halo component dominates. Previous studies commonly applied an ``aperture correction'' factor to scale the observed halo mass to the expected total. By analyzing several different cosmological hydrodynamical simulations \citep[][]{Bullock_2005,Rodriguez-Gomez_2016,Monachesi2019}, \citet{Bell_2017} derived a consistent average multiplicative correction factor of 3, scaling the halo mass measured within the minor-axis distance range of 10--40 kpc to the total halo mass for galaxies with stellar masses in the range of 3--10$\times$10$^{10}$ $M_{\odot}$. This factor carries a random scatter of about 40\% and a systematic uncertainty below 15\%. 

To apply this scaling factor to our sample, we first examine how best to normalize the minor-axis profiles of galaxies spanning a wide range of stellar masses and sizes. In Figure~\ref{Fig:Smb_norm}, we show the minor-axis surface brightness profiles of our galaxies and compare two normalization schemes. In the first, the minor-axis radius of each profile is normalized by its own major-axis half-light radius, $R_{\rm e, maj}$ (upper panels). In the second, the radius is normalized by the average $R_{\rm e, maj}$, denoted $<R_{\rm e, maj}>$, of galaxies of similar stellar mass (lower panels). $<R_{\rm e, maj}>$ is derived based on a broken log–log linear fit to the LOWESS-smoothed stellar mass-$R_{\rm e, maj}$ relation of our sample, parameterized as 
\begin{equation}
 \log<R_{\rm e, maj}> = k\cdot(\log M_{\star}/M_{\odot} - 9.5) + 0.66,
\end{equation}
where $k$ is 0.26 for $\log M_{\star}/M_{\odot} < 9.5$ and 0.10 for $\log M_{\star}/M_{\odot} \geq 9.5$. As shown in Figure~\ref{Fig:Smb_norm}, $<R_{\rm e, maj}>$-normalized minor-axis profiles exhibit significantly smaller scatter across the full sample compared to those normalized by each galaxy’s own $R_{\rm e, maj}$. The $<R_{\rm e, maj}>$-normalized minor-axis profiles of our galaxies display about self-similar shapes. 

For the stellar mass range analyzed by \cite{Bell_2017}, the minor-axis distance range of 10\,–\,40 kpc corresponds to $\sim$ 2 to 8 $\times$ $<R_{\rm e, maj}>$. We therefore estimate the total stellar halo mass in our galaxies by multiplying ``aperture mass'' measured within 
2 to 8 $\times$ $<R_{\rm e, maj}>$ by a factor of 3.

The stellar halo mass is derived from the $r$-band luminosity and its mass-to-light ratio (M/L), obtained using the M/L--($g-r$) relation from \cite{Zhang2017}. As shown by \cite{Zhang2017}, the M/L--($g-r$) relation is among those with the weakest dependence on star formation history (second only to $B-V$), and is therefore the optimal choice for estimating M/L of our stellar halos. This relation carries a random uncertainty of $\sim$0.2 dex under conditions of low dust extinction. The $(g\,-\,r)_{\rm outer}$ color used for M/L estimation is measured across stellar halo-dominated region of each galaxy. For a few galaxies with very low S/N measurement of colors, we use the M/L predicted by the best-fit $\log-\log$ linear relation between M/L and the stellar halo mass of the other galaxies.

\section{Results} \label{sec:results}

All basic galaxy properties and derived stellar halo measurements are provided in Tables~\ref{tab:sampletbl_basic} and~\ref{tab:sampletbl_SH} in the Appendix. Where applicable, reported fluxes and colors have been $K$-corrected to redshift $z=0$ following \cite{Chilingarian2010}, which presents analytic formulae based on observed colors and redshift.

\subsection{Stellar halo detection rate} \label{sec:detrate}
\begin{figure}[!ht]
  \centering
  \includegraphics[width=0.47\textwidth]{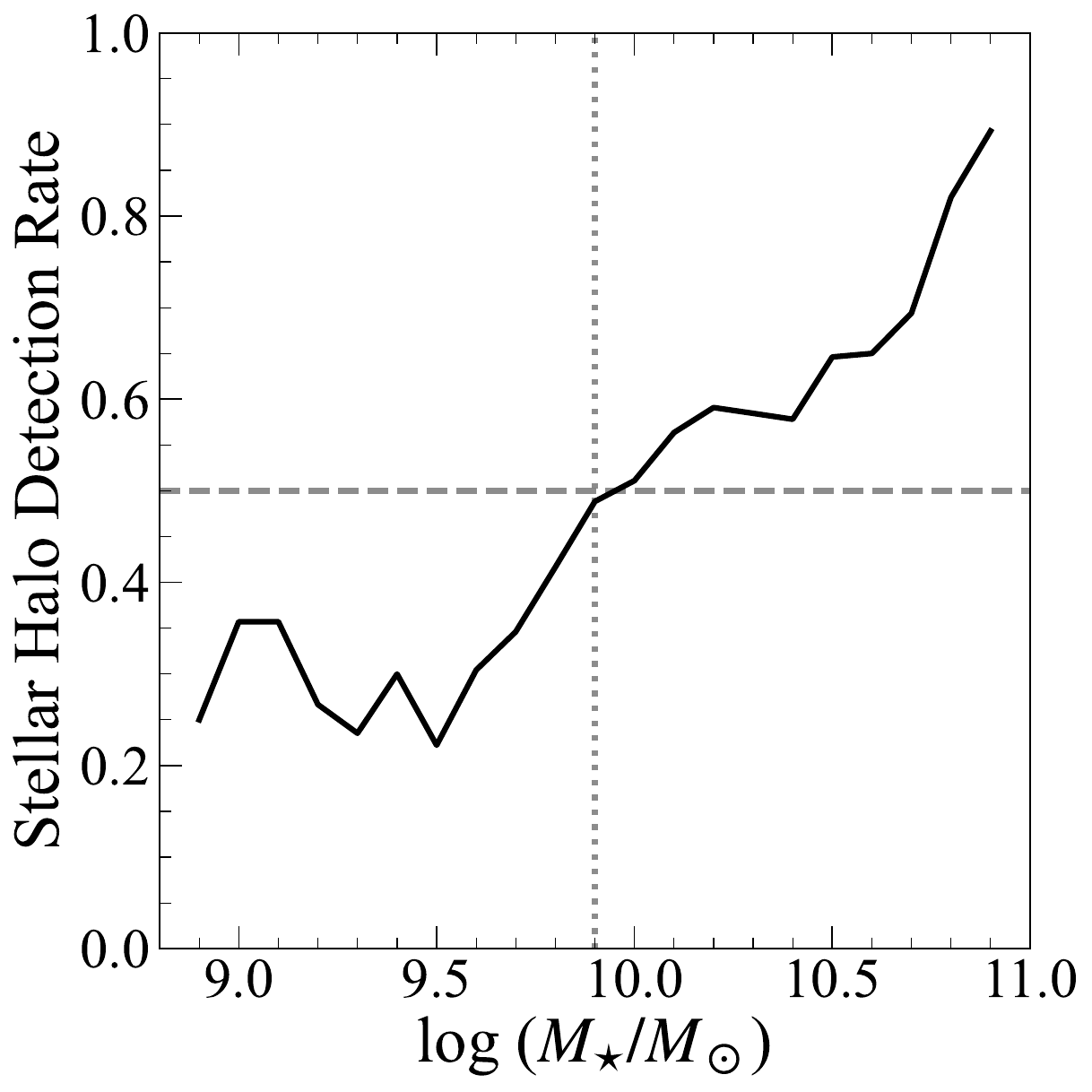} 
  \caption{Stellar halo detection rate as a function of galaxy stellar mass. The halo detection rate falls well below 50\% at $\log (M_{\star}/M_{\odot}) < 9.9$. See Section \ref{sec:detrate} for details.} \label{Fig:DetectionRate}
\end{figure}

It goes without saying that the ability to detect stellar halos is limited by the surface brightness depth of the imaging survey. As a result, halo detections are biased toward galaxies with relatively high halo fractions, with this bias becoming more pronounced at lower stellar masses. We quantify the halo detection rate as a function of galaxy stellar mass by adopting a 0.5 dex mass bin width and a 0.1 dex running step for the bin centers, evaluated from the high-mass end to the low-mass end of the sample. The resulting halo detection rate curve is shown in Fig.\ref{Fig:DetectionRate}. The stellar halo detection rate decreases from $\sim$90\% at the high-mass end to $\sim$50\% at stellar masses $\log (M_{\star}/M_{\odot}) \sim 9.9$, below which it declines rapidly and reaches a plateau of $\sim$30\% (with significant fluctuation) at $\log (M_{\star}/M_{\odot}) < 9.7$. A detection rate below 50\% makes it challenging to recover typical properties of the underlying population. We therefore restrict our subsequent statistical analyses to galaxies with $\log (M_{\star}/M_{\odot}) > 9.9$.

\subsection{Power-law indices of stellar halo profiles} \label{sec:Nindex}

\begin{figure*}[!ht]
  \begin{center}
    \includegraphics[width=1.0\textwidth]{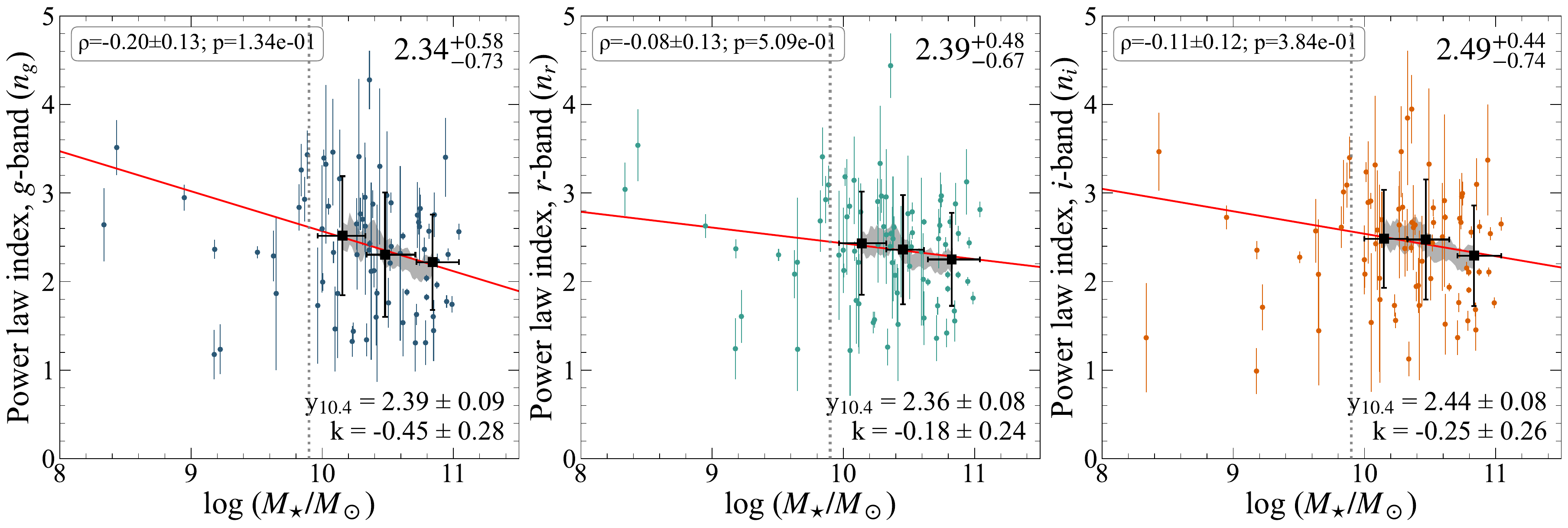}
    \caption{Power-law indices of stellar halo profiles are plotted against the host galaxy stellar mass. Results for the $g$-band profiles, $r$-band profiles and $i$-band profiles are respectively shown in the left, middle and right panels. The median and central 68\% range of the indices for galaxies with $\log M_{\star}/M_{\odot} > 9.9$ (vertical gray dotted lines) are listed in the top-right corners. The Spearman rank correlation coefficients, their uncertainties, and the p-values are listed in the top-left corners. Running means and the rms scatters are shown as grey-shaded region. Large black squares with error bars indicate the mean and central 68\% scatter of galaxies in three non-overlapping galaxy mass intervals. The best-fit slope ($k$) and intercept at $\log M_{\star}/M_{\odot}=10.4$ ($y_{10.4}$) are also provided, where $n=k\times(\log M_{\star}-10.4)+y_{10.4}$ and 10.4 is the median logarithmic stellar mass of galaxies with halo detection. See Section \ref{sec:Nindex} for details.}
    \label{Fig:Mass_N}
  \end{center}
\end{figure*}

Figure~\ref{Fig:Mass_N} presents the the power-law indices \textit{n} fitted to the minor-axis stellar halo profiles in the $g$, $r$, and $i$ bands as a function of the stellar mass for galaxies with well-sampled halo profiles (78 in $r$ band). While our sample includes galaxies with $\log (M_{\star}/M_{\odot})$ down to $\simeq$ 7.0, we focus on galaxies with $\log (M_{\star}/M_{\odot}) > 9.9$. 

The median and central 68\% scatter of the $n$ distributions for each band are reported in the upper-right corner of each panel. A weak but systematic dependence of the median $n$ on wavelength is observed, such that the median $n$ is larger in redder filters. This slight wavelength dependence indicates that stellar halo profiles are steeper at longer wavelengths, implying negative median color gradients in the stellar halos.

We perform a Spearman rank correlation test of the galaxy mass dependence of $n$, and find a weak ($-0.2 \leq \rho \leq -0.1$) negative correlation. We perform a linear fit to the relations. The best-fit slopes and intercepts are reported in each panel. The negative galaxy mass dependence of $n$ suggests that the stellar halos of more massive galaxies tend to have a more extended spatial distribution.

To place our findings in context, we compare the stellar halo profiles of our sample with those of the MW and other nearby galaxies. To make the comparison, we first convert the three-dimensional (3D) power-law indices reported in previous studies to projected values by subtracting 1. The projected power-law index of the MW reported in the literature spans a range of 1.5 to 2.5 \citep{Bell_2008, Juric_2008, Watkins_2009, Deason_2011, Sesar_2011, Pila_2015, Xue_2015, Das_2016, Ablimit_2018, Iorio_2018, Yang_2022, Chen_2023, Medina_2024}, which overlaps with the typical values of our sample. For M31, the power-law index is $\sim$2.6–2.7 \citep{Ibata_2014, Williams_2015}, in good agreement with the median of our sample. \citet{Harmsen_2017} reported power-law indices ranging from 2.0 to 3.7 for six nearby MW–mass galaxies from the GHOSTS survey, while \citet{Gilhuly_2022} found $n$ values between 1.7 and 3.3 for eleven out of twelve nearby edge-on disk galaxies. Overall, these literature measurements fall within the range spanned by our sample.

\subsection{Colors of stellar halos} \label{sec:haloclr}

\begin{figure*}[!ht]
  \begin{center}
    \includegraphics[width=1.0\textwidth]{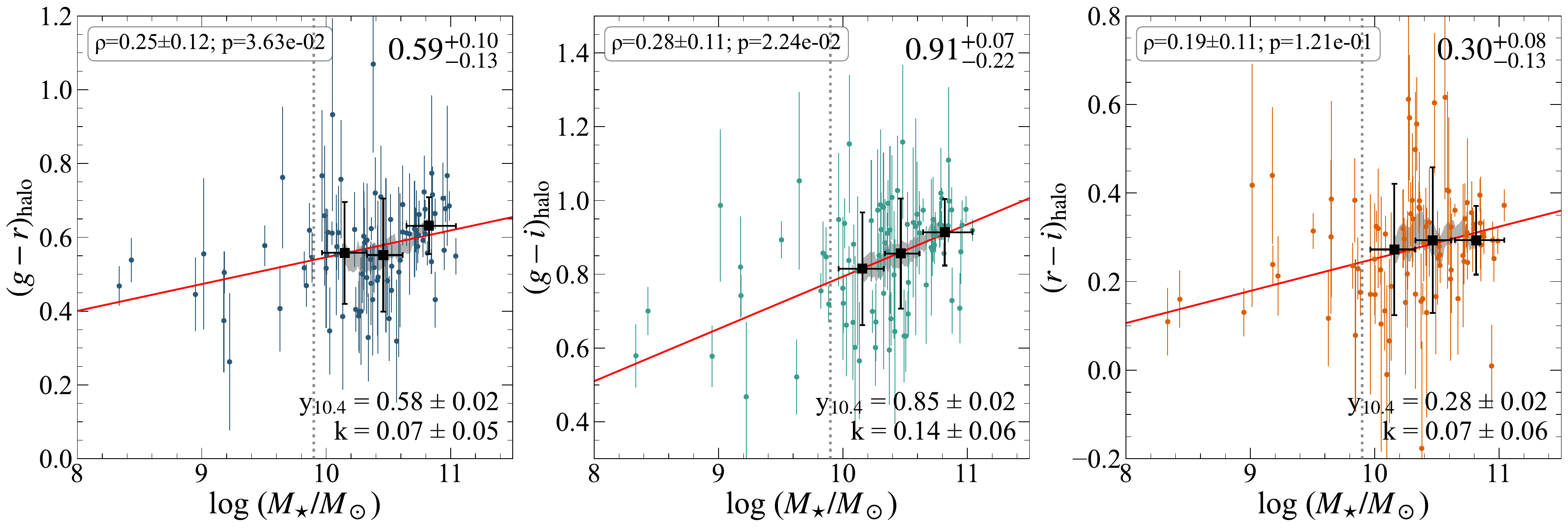}
    \caption{Similar to Figure \ref{Fig:Mass_N}, but showing the results of stellar halo colors of our galaxies. See Section \ref{sec:haloclr} for details.}
    \label{Fig:Mass_outclr}
  \end{center}
\end{figure*}

Figure~\ref{Fig:Mass_outclr} presents the integrated colors, $(g-r)_{\rm halo}$, $(g-i)_{\rm halo}$, and $(r-i)_{\rm halo}$ of the stellar halos as a function of the galaxy stellar mass. Nine galaxies with color signal-to-noise ratios less than 3 were not plotted.

The median and central 68\% scatter of the three colors are reported in the upper-right corner of each panel. The colors of $(g-r)_{\rm halo}$ and $(g-i)_{\rm halo}$ have a moderate ($\rho \sim 0.25$) and significant (p values $\leq$ 0.05) correlation with galaxy stellar mass, suggesting redder median colors of more massive galaxies. Linear least-squares fitting was performed for the color--mass distribution and the best-fit slopes and intercepts are reported in the figure.

\citet{Merritt_2016} reported $g-r$ colors ranging from 0.40 to 0.75 within 10--20 kpc along the major axis of six nearby galaxies from the Dragonfly Nearby Galaxies Survey (DNGS). \citet{Wang_2019} reported $g-r$ colors between 0.3 and 0.8 and $g-i$ colors between 0.6 and 1.2 within 10--20 kpc by stacking isolated central galaxies with total stellar masses of $10^{9.2}$--$10^{11.1}$\,\msun at redshift 0.05--0.1. They also found a clear positive stellar mass dependence of the colors. \citet{Gilhuly_2022} measured average outskirts $g-r$ colors of 0.40--0.75 for twelve nearby edge-on disk galaxies with stellar masses in the range $10^{9.68}$--$10^{10.88}$\,\msun. All of these previous color measurements occupy a color range similar to that found in our work. The redder typical colors of stellar halos in more massive galaxies reflect either older stellar ages or more importantly higher metallicities (see also Section \ref{sec:haloclrmass}).

\subsection{Inner boundary of the stellar halo-dominated region} \label{sec:smbhalo}

\begin{figure*}[!ht]
  \begin{center}
    \includegraphics[width=1.0\textwidth]{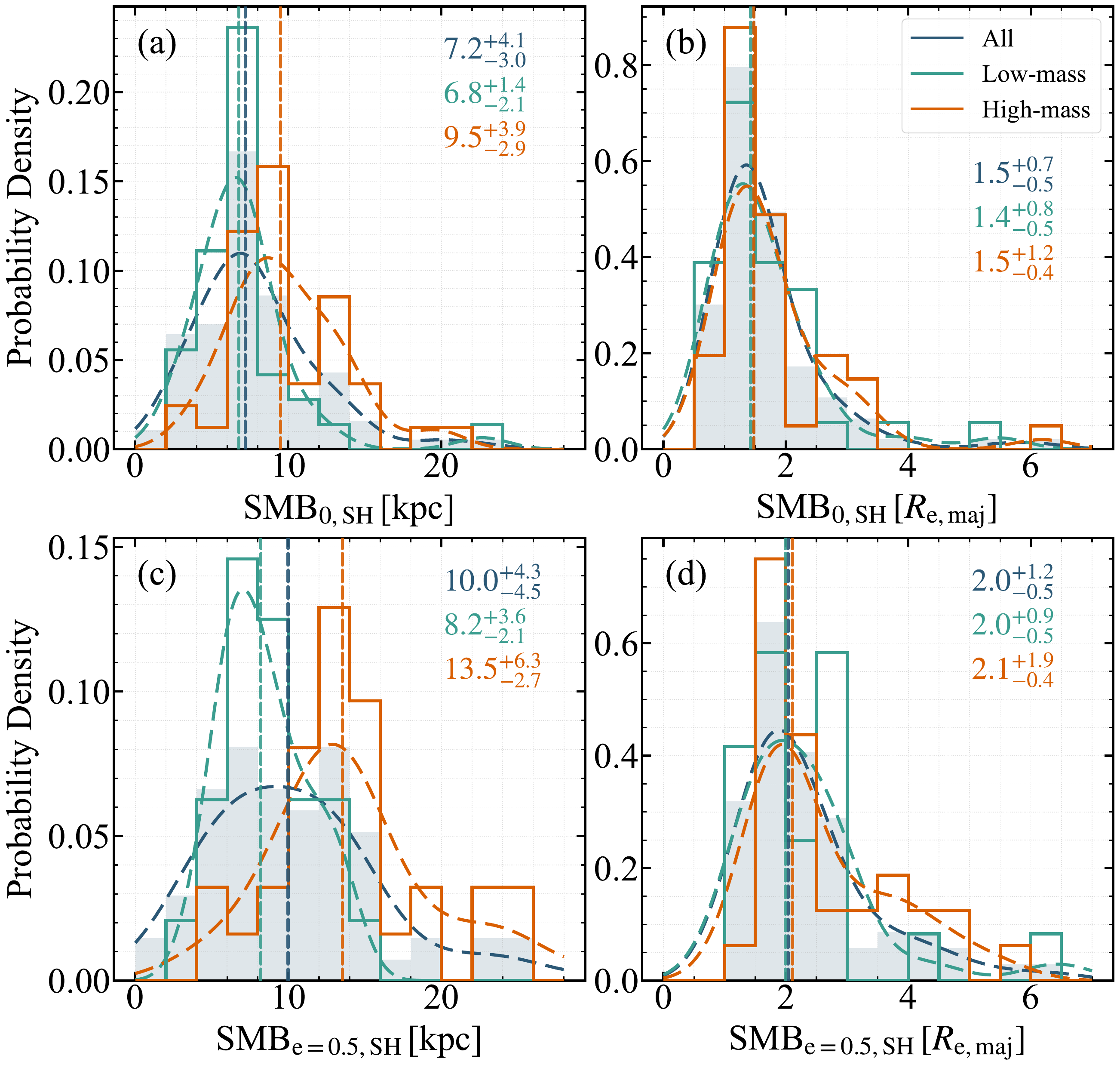}
    \caption{Distribution of the inner boundary of the stellar halo-dominated region of our galaxies. The boundary is defined as the semi-minor axis radius where stellar halo starts dominating over galaxy disk (SMB$_{\rm 0,\,SH}$; see Figure \ref{Fig:Measurement}).The boundary is expressed in units of kpc (panel a) or $R_{\rm e, maj}$ (panel b). The distribution for the most prominent stellar halos with ellipticity $< 0.5$ (SMB$_{\rm e=0.5,\,SH}$) is shown in units of kpc (panel c) or $R_{\rm e, maj}$ (panel d). The sample is divided into low- and high-mass subsamples at a stellar mass of $10^{10.4}$\,\msun. Colored vertical dashed lines indicate the median values of each subsample. The median and central 68\% range of the indices for full sample, low- and high-mass subsamples are given in each panel. Colored dashed curves show the corresponding kernel density estimation (KDE) profiles for each distribution, following the same color scheme as the histograms. See Section \ref{sec:smbhalo} for details.}
    \label{Fig:Hist_smbhalo}
  \end{center}
\end{figure*}

With a large sample of stellar halo detections in disk galaxies, it is helpful to examine the typical division radii between disk-dominated and halo-dominated regions. Besides the standard definition of inner boundaries described in Section \ref{sec:fullsample}, referred to as SMB$_{\rm 0,\,SH}$, here we also show inner boundary radii where the halo ellipticity first reaches 0.5 (Section \ref{sec:goldensample}), referred to as SMB$_{\rm e=0.5,\,SH}$. Figure \ref{Fig:Hist_smbhalo} presents the distributions of SMB$_{\rm 0,\,SH}$ in units of kpc (panel a) and $R_{\rm e, maj}$ (panel b), as well as SMB$_{\rm e=0.5,\,SH}$ in units of kpc (panel c) and $R_{\rm e, maj}$ (panel d). The effective radius $R_{\rm e, maj}$ is the semi-major axis radius of the isophote containing half of the total flux in the $r$ band for each galaxy.

In the left panels of Figure \ref{Fig:Hist_smbhalo}, the inner boundary measured in kiloparsecs shows a clear positive dependence on galaxy stellar mass, albeit with considerable overlap. The median values of SMB$_{\rm 0,\,SH}$ and SMB$_{\rm e=0.5,\,SH}$ are $9.5_{-2.9}^{+3.9}$ kpc and $13.5_{-2.7}^{+6.3}$ kpc for the high-mass subsample, whereas they are $6.8_{-2.1}^{+1.4}$ kpc and $8.2_{-2.1}^{+3.6}$ kpc, respectively, for the low-mass subsample.

In contrast, when expressed in units of $R_{\rm e, maj}$, the inner boundaries show virtually no dependence on galaxy mass. The median SMB$_{\rm 0,\,SH}$ and SMB$_{\rm e=0.5,\,SH}$ are $1.5_{-0.4}^{+1.2}\,R_{\rm e, maj}$ and $2.1_{-0.4}^{+1.9}\,R_{\rm e, maj}$ for the high-mass subsample, very close to $1.4_{-0.5}^{+0.8}\,R_{\rm e, maj}$ and $2.0_{-0.5}^{+0.9}\,R_{\rm e, maj}$ for the low-mass subsample. The corresponding median values for the full sample are $1.5_{-0.5}^{+0.7}\,R_{\rm e, maj}$ and $2.0_{-0.5}^{+1.2}\,R_{\rm e, maj}$, respectively.

These results suggest that a fixed inner boundary in kiloparsecs cannot be used to measure stellar halos across a wide stellar mass range. Instead, a fixed minor-axis radius of $\sim$1.5--2.0$R_{\rm e, maj}$, corresponding to $\sim$3--4$R_{\rm e, maj}$ along the major axis for an axis ratio of 0.5, provides a more reasonable choice for statistical studies of large samples of high-inclination galaxies. Despite the constant median values in units of $R_{\rm e, maj}$, the distribution of inner boundaries has a tail extending to $\sim$ 6$R_{\rm e, maj}$. Our results justify the conservative choice of \citet{Gilhuly_2022}, who adopted five times the half-mass radius along the major axis to identify stellar halos. 

\subsection{Stellar halo color--mass relations} \label{sec:haloclrmass}

\begin{figure*}[!ht]
  \begin{center}
    \includegraphics[width=1.0\textwidth]{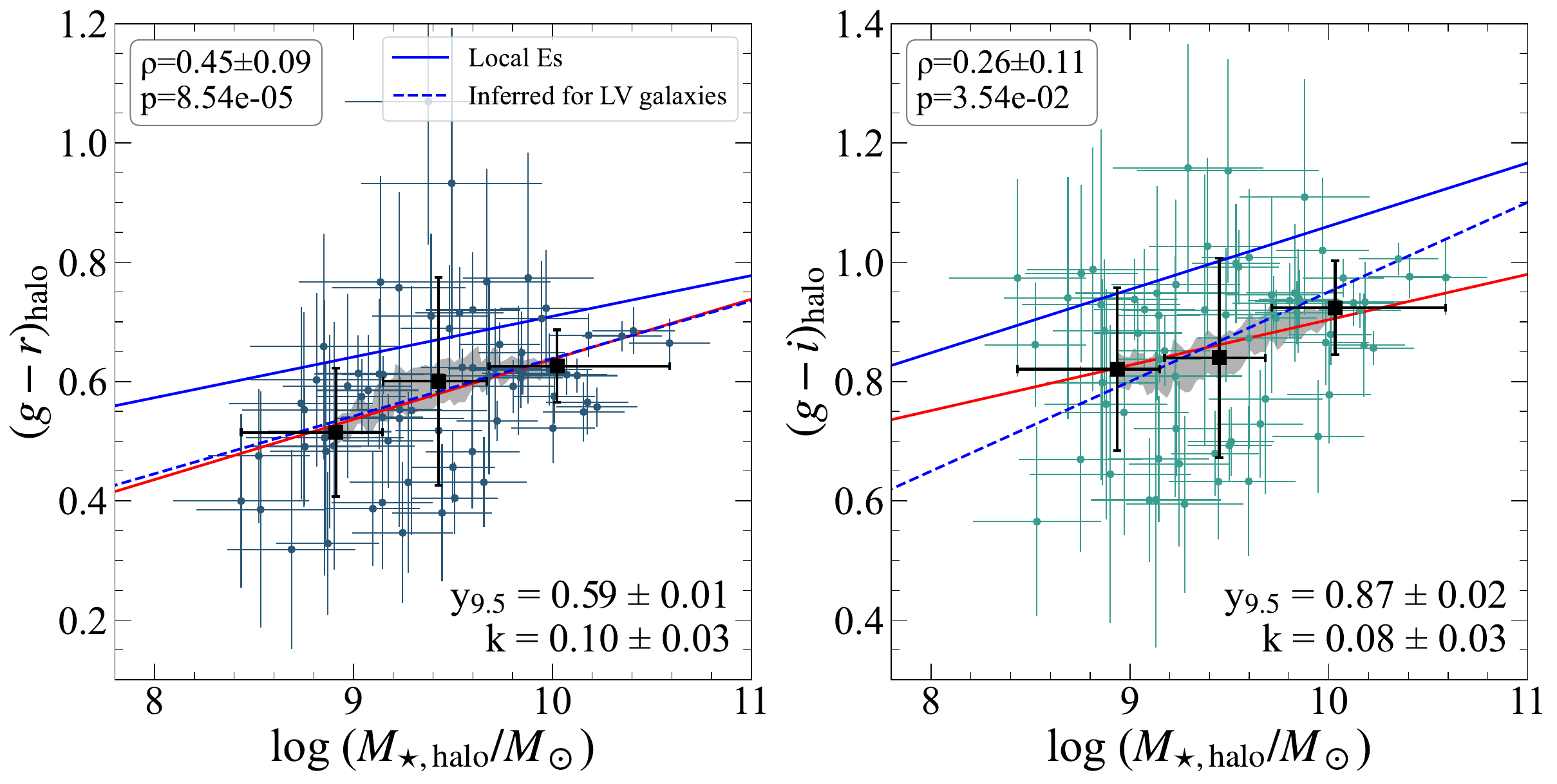}
    \caption{Stellar halo colors are plotted against stellar halo masses for our galaxies. Results for the $(g-r)_{halo}$ and $(g-i)_{halo}$ are respectively shown in the left and right panels. The blue solid lines show the color--mass relation of local elliptical galaxies, while the blue dashed lines represent the relation inferred from the halo mass--metallicity correlation of the stellar halos of 14 Local Volume (LV) MW-mass galaxies \citep{Smercina_2022}. The Spearman rank correlation coefficients, their uncertainties, and the p-values are listed in the top-left corners. Running average and the rms scatter are shown as grey-shaded region. Large black squares with error bars indicate the average and central 68\% scatter of galaxies in three separate galaxy mass intervals. The best-fit linear relations (color = $k\times(\log M_{\star}-9.5)+y_{9.5}$) for our sample are shown as red lines, and the best-fit slope ($k$) and intercept at $\log M_{\star}/M_{\odot}=9.5$ ($y_{9.5}$) are also provided. See Section \ref{sec:haloclrmass} for details.}
    \label{Fig:SH_color_mass}
  \end{center}
\end{figure*}

Figure~\ref{Fig:SH_color_mass} presents the integrated colors, $(g-r)_{\rm halo}$ and $(g-i)_{\rm halo}$ of the stellar halos as a function of the stellar halo mass for our sample. Nine galaxies with color uncertainties larger than 0.33 mag are not plotted. As is shown, the stellar halo colors have a moderate ($\rho \sim 0.3-0.5$) yet significant (p values $<$ 0.05) correlation with the mass. 

Quiescent galaxies follow a tight color--magnitude or color--mass relation (commonly referred to as the red sequence), which is primarily driven by the correlation between metallicity and galaxy mass, whereas stellar age shows a weaker dependence on luminosity or stellar mass and mainly contributes to the color scatter at fixed mass \citep[e.g.,][]{Graves2009}. Since stellar halos generally consist of old stellar populations \citep[e.g.,][]{Bressan1996, Carollo_2016}, the observed stellar halo color--mass correlation is likely mainly driven by a stellar halo mass--metallicity relation.

\citet{Smercina_2022} observed a stellar halo mass--metallicity relation for 14 MW–mass galaxies in the Local Volume. To enable a direct comparison with our measurements, we convert the \citet{Smercina_2022} mass--metallicity relation into color--mass relations using empirical color--metallicity correlations. Specifically, we derive a relation between color and metallicity by combining the galaxy stellar mass--metallicity relations from the local universe \citep[][]{Gallazzi2005, Kirby2013} with the stellar mass--color relation of local elliptical galaxies \citep{Chang2006}\footnote{We apply $K$-corrections to the stellar color–mass relations reported in the paper, shifting them to $z=0$.}. The resulting stellar halo ($g-r$)--mass and ($g-i$)--mass relations for the 14 MW–mass galaxies are overplotted as dashed lines in Fig.~\ref{Fig:SH_color_mass}. The inferred ($g-r$)--mass relation agrees well with our measurements, whereas the inferred ($g-i$)--mass relation shows a slight systematic offset. The slight offset for ($g-i$)--mass relation may be explained by its higher sensivity to stellar ages.

The presence of a stellar halo color--mass correlation, and its overall consistency with that inferred from the mass–-metallicity relation, confirms the existence of a general halo mass–-metallicity correlation. This correlation implies that stellar halos are predominantly built from the remnants of one or a few massive satellites, rather than from the accretion of numerous low-mass ones. This interpretation is consistent with theoretical expectations \citep[e.g.,][]{DSouza_2018, Monachesi2019}.

\subsection{Stellar halo fraction} \label{sec:fhalo}

\begin{figure*}[!ht]
  \begin{center}
    \includegraphics[width=0.95\textwidth]{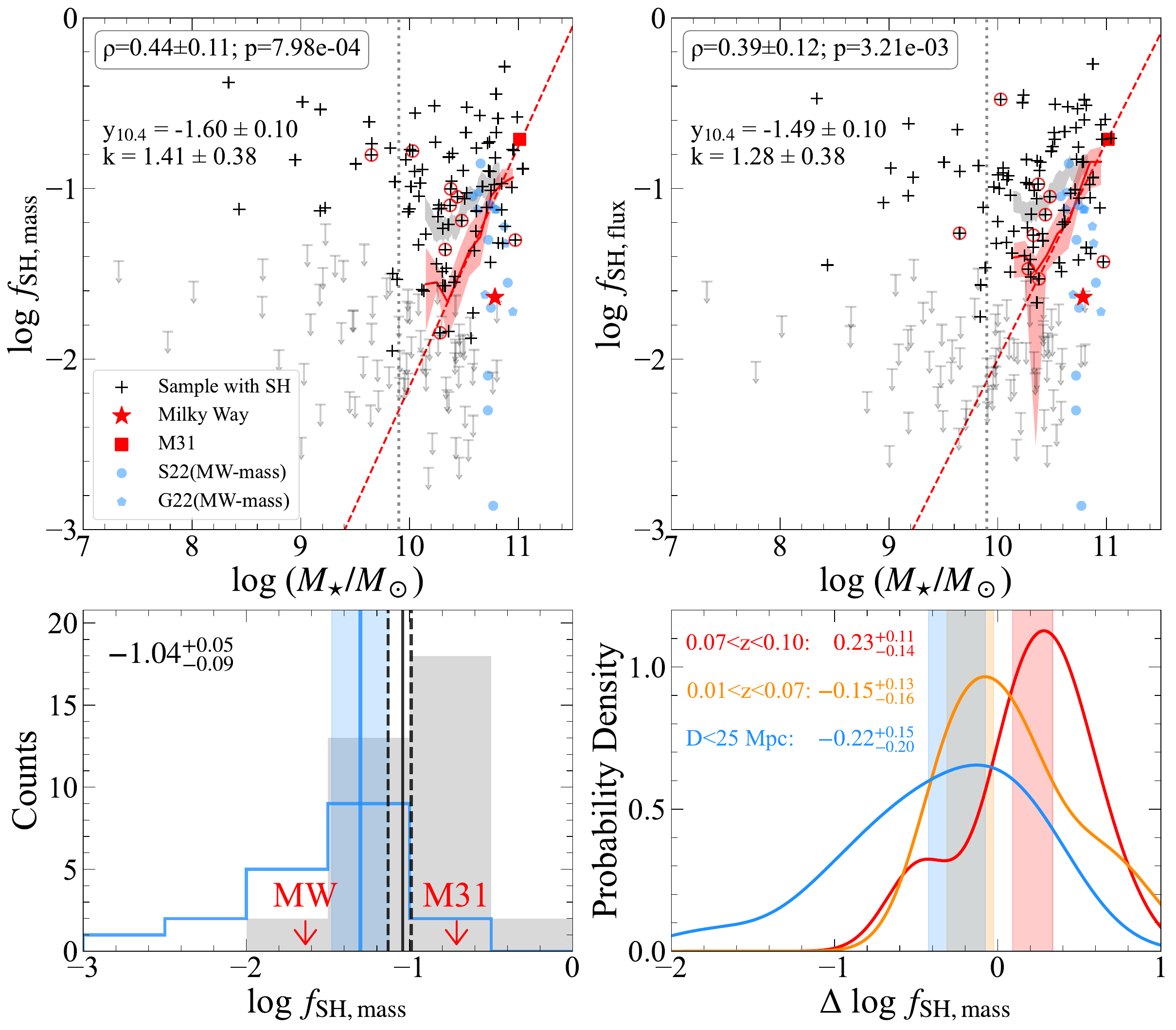}
    \caption{Stellar halo mass and flux fractions relative to host galaxies. \textbf{Upper Left \& Right:} stellar halo mass/flux fraction vs. galaxy stellar mass. Black plus symbols denote galaxies with halo detections; those enclosed by red circles have power-law surface profiles and $q < 0.4$ in the outskirts. Gray arrows indicate upper limits for non-detections. The gray-shaded region shows the rms scatter of mean fractions from 1000 bootstrap resamplings of detected halos in running stellar-mass bins. Literature measurements for the MW (red pentagram) and M31 (red square), and 17 nearby MW–mass galaxies ($10.5 < \log M_{\star}/M_{\odot} < 11.0$) within 25 Mpc from \cite[][S22]{Smercina_2022} and \cite[][G22]{Gilhuly_2022} (dodger blue), are included. Spearman coefficients, uncertainties, and $p$-values for detected halos at $\log M_{\star}/M_{\odot} > 10.3$ are listed. The red solid line and shaded region show the Kaplan–Meier (KM) estimator median halo fractions and $1\sigma$ scatter over 0.5-dex running mass bins with 0.1-dex steps. Best-fit \texttt{LINMIX} relations for $\log M_{\star}/M_{\odot} \geq 10.2$, $\log f_{\rm SH} = k(\log M_{\star}-10.4)+y_{10.4}$, are shown as red dashed lines, with $k$ and $y_{10.4}$ listed. \textbf{Lower Left:} histograms of stellar halo mass fractions of MW-mass galaxies. Gray-shaded and dodger-blue open histograms respectively represent galaxies from our sample and nearby galaxies within 25 Mpc. Vertical solid and dashed lines mark the median (based on KM estimator for our sample) and the central 68\% intervals. Dodger-blue-shaded region represents the 68\% confidence intervals from 1000 bootstrap resamplings of nearby galaxies; \textbf{Lower Right:} kernel density estimates of differential stellar halo fraction distributions for MW-mass galaxies within $<25$ Mpc and MW-mass galaxies below and above $z=0.07$ in our sample. The differential halo mass fractions are derived by subtracting the median stellar halo mass fraction--galaxy mass relation from individual galaxies. Median values and 68\% confidence intervals are shown by color-shaded regions and listed. See Section~\ref{sec:fhalo} for details.}
    \label{Fig:SH_fraction}
  \end{center}
\end{figure*}

\subsubsection{The galaxy mass dependence for our sample}\label{sec:fhalo_our}

The flux or mass fraction of stellar halos relative to the total stellar mass of a galaxy reflects the cumulative contribution of merger events in the past. In the upper panels of Fig.\ref{Fig:SH_fraction}, we present the dependence of stellar halo mass and $r$-band flux fractions on galaxy stellar mass.

To explore the dependence of stellar halo fractions on galaxy stellar mass, it is necessary to account for both galaxies with and without stellar halo detections. We therefore estimate upper limits on the stellar halo fractions for galaxies without detected halos. Given that such galaxies are dominated by disk light out to the surface brightness limit of $\simeq$ 30 \magsqarc, the stellar halo component is expected to contribute no more than half of the observed stellar light at this surface brightness level. This corresponds to a maximum stellar halo surface brightness of 30.75 \magsqarc\ at the outermost galactocentric radii  where galaxy light is detected.

The above upper limit, together with the galaxy-mass-dependent median power-law indices and mass-to-light ratios derived from galaxies with halo detections, allows us to place an upper limit on the total stellar halo mass for galaxies without detections. With both detections and upper limits included for the full sample, we employ two different methods to estimate the dependence of stellar halo fraction on galaxy stellar mass. The first method is the Kaplan–Meier (KM) estimator, a commonly used survival analysis technique in astronomy for estimating median or mean values for data sets that include non-detections \citep[e.g.,][]{Calette2018}. The second method is the Bayesian linear regression approach \texttt{LINMIX} \citep{Kelly2007}, which performs linear regression while explicitly accounting for non-detections.

The median stellar halo fractions and their 68\% percentile confidence intervals, derived from the KM estimator over 0.5-dex running windows in galaxy stellar mass, together with the maximum-likelihood linear relation obtained from \texttt{LINMIX}, are overplotted in the upper panels of Fig.~\ref{Fig:SH_fraction}. The most likely \texttt{LINMIX} slope and intercept are also listed in each panel. We note that the \texttt{LINMIX} linear regression is restricted to galaxies with $\log (M_{\star}/M_{\odot}) > 10.2$, above which the KM estimator yields a virtually $\log-\log$ linear relationship (see the red shaded regions in Fig.~\ref{Fig:SH_fraction}). Reassuringly, the two methods yield highly consistent median trend for the dependence of stellar halo fraction on galaxy stellar mass.

The median stellar halo flux fraction decreases monotonically with decreasing galaxy mass above the 50\% detection rate limit, albeit with a larger uncertainty at $\log (M_{\star}/M_{\odot}) \lesssim 10.3$ (where detection rate is $\lesssim$ 60\%). At $\log (M_{\star}/M_{\odot}) < 10.2$, all galaxies exhibit stellar halo fractions higher than that expected from the linear extrapolation of the relation at higher masses (red dashed lines in the upper panels of Fig.~\ref{Fig:SH_fraction}), with no clear dependence on galaxy mass for either detections or upper limits. This flattening trend at the low mass end likely reflects a detection bias due to the low detection rate, but may also partly indicate an intrinsically weak mass dependence, as predicted by cosmological simulations \citep[e.g.,][]{Rodriguez-Gomez_2016, Shi2022}. 

To be quantitative, there is a strong correlation ($\rho \sim 0.44$) between the halo mass fraction and the galaxy stellar masses at $\log M_{\star}/M_{\odot} > 10.3$. The correlation coefficient for stellar halo flux fraction is slightly lower ($\rho \sim 0.39$).

The lower-left panel of Fig.~\ref{Fig:SH_fraction} shows the histogram of the stellar halo mass fraction for MW–mass galaxies with $10.5 < \log(M_\star) < 11.0$. The KM estimator median stellar halo mass fraction (in logarithm) for this subsample is $-1.0$ dex (with a $+$0.2/$-$0.3 dex uncertainty considering the 95\% confidence intervals), $\sim$ 0.1 dex lower than the median for galaxies with halo detections. There is a substantial spread in the halo fractions, with the fifth–95th percentiles spanning $-1.5$ to $-0.5$ dex for galaxies with detected halos, indicating diverse merger histories. 
We note that our results for MW-mass galaxies in this paper are virtually identical when stellar halos are measured over a fixed minor-axis radial range of 10--40 kpc, rather than a normalized radial range.

\subsubsection{Comparison with MW and other nearby galaxies}\label{sec:fhalo_comp}

With the first highly complete stellar halo detections for disk galaxies at $\log(M_\star) \gtrsim 10.0$, we can assess how representative familiar systems such as the MW and M31 are in terms of their stellar halo fractions. 

The most up-to-date estimate of the logarithmic stellar halo mass fraction of the MW is $-1.64\pm 0.15$ for a Kroupa IMF \citep{Deason_2019}, about 0.43 dex larger than the value reported earlier by \citet{Harmsen_2017}. As shown in Fig.~\ref{Fig:SH_fraction}, the MW’s stellar halo mass fraction is $\sim$ 0.6 dex lower than the typical value for galaxies of similar mass, suggesting an unusually quiescent merger history. We note that a relatively quiet merger history for the MW has been inferred by several recent studies \citep{Ruchti2015, Fragkoudi2020, Deason2023}. In contrast to the MW, the stellar halo mass fraction of M31 \citep[$-0.71\pm 0.30$ dex;][]{Sick_2015, DSouza_2018_NatAs, Smercina_2022} follows the median halo fraction--galaxy mass relation.

The largest samples of stellar halo measurements of nearby galaxies were presented by \cite{Smercina_2022} and \cite{Gilhuly_2022}. Together, the two studies include 17 unique extragalactic MW-mass galaxies ($10.5 < \log M_{\star}/M_{\odot} < 11.0$) within 25 Mpc. For the galaxies NGC 4565 and M101, which are common to both studies, we adopt the halo measurement from \cite{Smercina_2022}. \cite{Smercina_2022} measured stellar halo mass as three times of the mass enclosed within a minor-axis radial range of 10--40 kpc, whereas \cite{Gilhuly_2022} reported halo fractions measured beyond 20\,kpc along major axis which is approximately equivalent to a minor-axis radius of 10 kpc \citep[see][for a comparison]{Gilhuly_2022}. We note that the surface brightness profiles of eight galaxies studied by \cite{Gilhuly_2022} have outer minor-axis radial boundaries of less than 30 kpc, which fall well short of the 40 kpc requirement. These eight galaxies are therefore excluded from our comparison.  After applying the same multiplicative factor of three to the \cite{Gilhuly_2022} measurements for consistency, the resulting halo fractions for all 17 galaxies are overplotted in the upper panels of Fig.~\ref{Fig:SH_fraction}. 

Interestingly, 68\% (13/19) of the nearby galaxies, including the MW and M31, lie below the median relation defined by our sample (upper panels of Fig.~\ref{Fig:SH_fraction}), and this fraction is 72\% when considering only the 11 Local Volume galaxies. The median stellar halo fraction of these nearby galaxies ($-$1.30 dex, or $-$1.55 dex when considering only Local Volume galaxies) is $\sim$ 0.3 dex lower than the median value of our MW–mass galaxies (lower-left panel of Fig.~\ref{Fig:SH_fraction}).

\subsubsection{Dependence on heliocentric distance}\label{sec:fhalo_dist}

In light of the significantly lower typical stellar halo fractions of nearby MW-mass galaxies found above, we explore potential systematic variations of the halo fractions with heliocentric distance. To this end, we divide our full MW-mass sample ($\log M_{\star}/M_{\odot} \geq 10.5$, including both detections and upper limits) into two halves at a redshift of 0.07 and compare these two redshift intervals with the nearby galaxy sample at distances $<25$ Mpc. We emphasize that the galaxy sample at $\log M_{\star}/M_{\odot} \geq 10.5$ is complete up to the redshift limit of 0.1 in our analysis (Figure \ref{Fig:Z-Mass}), ensuring a meaningful comparative analysis of subsamples across different redshift intervals.

Because there is a significant halo fraction--galaxy mass correlation, we subtract (in logarithmic space) the halo fraction predicted by the median relation (the red dashed lines in the upper panels of Fig.~\ref{Fig:SH_fraction}) from the measured value for each galaxy in the three samples introduced above. This subtraction is necessary to avoid potential biases introduced by differences in the galaxy mass distributions among the three samples. The resulting distributions of the differential halo fraction ($\Delta\log f_{\rm SH, mass}$) of the three samples are shown in the lower-right panel of Fig.~\ref{Fig:SH_fraction}. We use kernel density estimates to represent the distributions. The median $\Delta\log f_{\rm SH, mass}$ and its 68\% confidence interval are indicated as color-filled regions and listed in the top-left corner of the panel. The values for our two samples are derived from the KM estimator.

As shown, there is a systematic increase in the median halo fractions with heliocentric distance, although the difference between that of the nearby galaxies and our lower-redshift sample is not statistically significant. It is worth noting that the median stellar halo fraction of our $z>0.07$ galaxies, i.e., $0.23 + (-1.04) = -0.81$ dex, where $-1.04$ is the KM estimator median fraction of the full MW-mass sample (lower-left panel of Fig.~\ref{Fig:SH_fraction}), is in good agreement with the median accreted stellar fraction ($-0.72$ dex) of MW-mass galaxies in general in the TNG50 simulations \citep{SotilloRamos2022}. This implies unusually quiescent merger histories for galaxies in the lower-redshift samples.

It has been found that the local universe out to redshift 0.07 ($\sim$ 300 Mpc) is under-dense relative to the large-scale cosmic average, with the matter density gradually increasing with distance within this volume, beyond which it remains more or less consistent with the cosmic average \citep[e.g.,][]{Keenan2013, Banik2025}. Together, these results help clarify the conundrum identified by the Dragonfly team \citep{Merritt2020} for nearby galaxies, where typical stellar halo fractions appear lower than predicted by cosmological hydrodynamical simulations. These findings also align with \cite{Wang2021}, who reported that MW–mass galaxies within $\lesssim$ 40 Mpc exhibit an anomalous deficit of low-mass satellite galaxies relative to averages measured over larger scales.

\section{Summary} \label{sec:summary}

We utilize deep imaging data from the Hyper Suprime-Cam Subaru Strategic Program Survey Public Data Release 3 (HSC-SSP PDR3) Deep/UltraDeep layers to investigate and characterize the stellar halo properties of 169 high-inclination central galaxies with stellar masses ranging from $10^{7.33}$ to $10^{11.04}$\msun. The central galaxy sample is constructed from the SDSS DR7 and GAMA DR4 spectroscopic surveys, selecting galaxies with $r$-band magnitudes brighter than 17.77 mag at $z < 0.1$. High-inclination systems are selected by requiring deconvolved disk axis ratios $q_{\rm disk} < 0.4$.

As dynamically hot structures, stellar halos are generally more spherical than disks. Our primary criterion for stellar halo identification (Section \ref{sec:fullsample}), applied to images corrected for PSF-induced scattered light, is based on isophotal ellipticity (or equivalently, axis ratios). This represents an improvement over previous approaches that relied on somewhat arbitrary galactocentric radii or multi-component surface brightness decompositions without clear justification. Using this method, we detect stellar halos in 93 galaxies.

This is the first systematic census of stellar halos in a flux-limited, large sample of high-inclination central disk galaxies in the local universe. The detection rate exceeds 50\% for galaxies with stellar masses above $10^{9.9} M_{\odot}$ and rises to $\gtrsim$ 70\% for MW-mass galaxies (Figure \ref{Fig:DetectionRate}). This sample will be used to put unique constraints on the accretion histories of disk galaxies and their connection to other galaxy properties in an accompaning paper. In the current paper, we provide the basic galaxy properties (Tables~\ref{tab:sampletbl_basic}) and derive the stellar halo measurements of the sample, including power-law surface brightness profile fitting, colors, halo mass and fractions (Tables~\ref{tab:sampletbl_SH} in Appendix~\ref{sec:proptbl}). The main results of this work are as follows:

\begin{enumerate}
    \item By analyzing a subsample of the 49 most prominent stellar halos (Section \ref{sec:goldensample}), we found that five of them are best described by an exponential radial decline in surface brightness, 44 by a single or broken power-law model. This result demonstrates that stellar halos generally follow power-law surface brightness profiles. 
    
    \item The power-law indices $n$ of stellar halo surface brightness profiles exhibit weak but systematic variation from the $g$ to $i$ band, with the redder bands having larger median $n$ and thus steeper radial decline, implying negative color gradients of stellar halos. In addition, more massive galaxies have lower median $n$ values, indicating their more extended stellar halo profiles (Section \ref{sec:Nindex}).
    
    \item The optical colors of the stellar halos exhibit a positive dependence on stellar mass of the host galaxies, with more massive galaxies having redder median colors (Section \ref{sec:haloclr}). The redder colors likely arise from a generally higher metallicity of stellar halos in more massive galaxies.
    
    \item The inner radial boundary, where stellar halos begin to dominate over stellar disks at larger minor-axis radii, shows a significant positive dependence on galaxy stellar mass when expressed in kiloparsecs. However, this systematic dependence largely disappears when the inner boundaries are normalized by the major-axis half-light radius $R_{\rm e, maj}$, yielding a nearly constant median value of $\sim 1.5-2.0R_{\rm e, maj}$ (depending on the exact definition) along the minor axis (Section \ref{sec:smbhalo}).

        \item The optical colors of stellar halos show a moderately strong positive correlation with stellar halo mass. This correlation primarily reflects an underlying stellar halo mass--metallicity relation, indicating that stellar halos are largely built from the remnants of a few massive (rather than numerous very low-mass) merger events, consistent with cosmological simulations.
        
    \item There exists a significant positive correlation between stellar halo fractions and galaxy stellar mass for galaxies with $\log (M_{\star}/M_{\odot}) \gtrsim 10.0$, where the halo detection rate is relatively high (Figure \ref{Fig:SH_fraction}). This is consistent with theoretical expectations that more massive galaxies generally experienced more active merger histories. 
    
    \item Our MW has a stellar halo fraction $\sim$ 0.6 dex below the median stellar halo fraction--galaxy mass relation. Comparing the stellar halo fraction distributions of nearby galaxies within 25 Mpc with those of the lower-redshift ($z < 0.07$) and higher-redshift galaxies in our sample indicates a systematic increase in typical halo fraction with heliocentric distance, with the galaxies at $z > 0.07$ showing median values in good agreement with IllustrisTNG predictions. The unusually quiescent merger histories implied for nearby galaxies appear to align with the under-dense nature of the local universe out to $z \lesssim 0.07$ (Section~\ref{sec:fhalo_dist}).

\end{enumerate}

\begin{acknowledgments}
\small
This work has been supported by the National Key Research and Development Program of China (No. 2023YFA1608100), the NSFC (Nos. 12122303, 11973039), and the China Manned Space Program with grant Nos.CMS-CSST-2021-B02, CMS-CSST-2021-A07. 
We also acknowledge support from the CAS Pioneer Hundred Talents Program, and the Cyrus Chun Ying Tang Foundations. 
HYW is supported by the National Natural Science Foundation of China (NSFC, Nos. 12192224) and CAS Project for Young Scientists in Basic Research, Grant No. YSBR-062.
GC acknowledges support provided by the Spanish Ministerio de Ciencia, Innovación y Universidades (MICIU) through the project PID2023-153342NB-I00 / 10.13039/501100011033

The Hyper Suprime-Cam (HSC) collaboration includes the astronomical communities of Japan and Taiwan,and Princeton University. The HSC instrumentation and software were developed by the National Astronomical Observatory of Japan (NAOJ), the Kavli Institute for the Physics and Mathematics of the Universe (Kavli IPMU), the University of Tokyo, the High Energy Accelerator Research Organization (KEK), the Academia Sinica Institute for Astronomy and Astrophysics in Taiwan (ASIAA), and Princeton University. Funding was contributed by the FIRST program from the Japanese Cabinet Office, the Ministry of Education, Culture, Sports, Science and Technology (MEXT), the Japan Society for the Promotion of Science (JSPS), Japan Science and Technology Agency (JST), the Toray Science Foundation, NAOJ, Kavli IPMU, KEK, ASIAA, and Princeton University.

This work is based in part on data collected at the Subaru Telescope and retrieved from the HSC data archive system, which is operated by Subaru Telescope and Astronomy Data Center at National Astronomical Observatory of Japan.
\end{acknowledgments}

\appendix

\section{Estimate of local background and its noise}\label{sec:backnoise}

We define a series of concentric annular regions centered on each of our sample galaxies, with the radius ratio between adjacent annulus set to 1.1. The intensity of each annulus is defined as the median of unmasked pixels within the annulus. The background region is defined as the first occurrence of six contiguous annuli where the radial intensity gradient is close to zero. We then divided the background region into 24 quadrants along the major and minor axis of these central galaxies and utilized the \texttt{sigma\_clipped\_stats} function to measure the 5-sigma clipped median ($I_{\rm sky,\,i}$) and standard deviation ($\sigma_{\rm sky,\,i}$) of unmasked pixels in i-th quadrant. The local background value was set as the median of $I_{\rm sky,\,i}$. The high-frequency background noise $\Delta I_{\rm sky,\,high}$ was calculated as the median of $\sigma_{\rm sky,\,i}/{\sqrt{N_{\rm sky,\,i}}}$, where $N_{\rm sky,\,i}$ indicates the number of unmasked pixels in a given quadrant. The low-frequency background noise $\Delta I_{\rm sky,\,low}$ was calculated by combining the standard deviation of the $I_{\rm sky,\,i}$ and the high-frequency background noise, see \cite{Gil_2005} for details. In addition to the background noise $\Delta I_{\rm sky}$, the uncertainty of the median background value is derived as $\Delta I_{sky}/\sqrt{24}$. By doing the above, we obtained an estimate of local depth of surface photometry for each galaxy as $2\,\times\,\Delta I_{\rm sky}$. 

\bibliography{sh}{}
\bibliographystyle{aasjournal}

\section{Foundational data products of this work}\label{sec:proptbl}

The appendix provides Tables~\ref{tab:sampletbl_basic} and~\ref{tab:sampletbl_SH}, which summarize the core measurements of the high-inclination central galaxies analyzed in this work. Table~\ref{tab:sampletbl_basic} lists the basic galaxy properties, and Table~\ref{tab:sampletbl_SH} presents the derived stellar halo parameters. The galaxies are grouped by stellar halo detection status and ordered by galaxy stellar mass. All measurements are based on imaging data corrected for PSF-induced scattered light. Colors and stellar masses are $K$-corrected and corrected for Galactic extinction. Reported uncertainties correspond to the 1$\sigma$ confidence intervals.

\clearpage{}
\centerwidetable
\begin{deluxetable*}{l r r r r r r r r r}
\setlength{\tabcolsep}{4pt}
\tabletypesize{\normalsize}
\tablecaption{Basic properties of high-inclination central galaxies}
\label{tab:sampletbl_basic}
\tablehead{
\colhead{Name} &
\colhead{RA(J2000)} &
\colhead{Decl(J2000)} &
\colhead{redshift} &
\colhead{$\log(M_{\rm \star})$} &
\colhead{$m_{\rm r}$} &
\colhead{$R_{\rm e}$} &
\colhead{$C_{\rm 28}$} &
\colhead{$f_{\rm bulge}$} &
\colhead{$\epsilon_{\rm outer}$} \\
\colhead{} & \colhead{(degree)} & \colhead{(degree)} & \colhead{} & \colhead{[\msun]} & \colhead{} & \colhead{[kpc]} & \colhead{} & \colhead{} & \colhead{} \\
\colhead{(1)} &
\colhead{(2)} &
\colhead{(3)} &
\colhead{(4)} &
\colhead{(5)} &
\colhead{(6)} &
\colhead{(7)} &
\colhead{(8)} &
\colhead{(9)} &
\colhead{(10)}
}
\startdata
\hline
\multicolumn{10}{c}{\textbf{Galaxies with stellar halo detection}} \\
\hline
VAGC-13467 & 151.05717 & +0.81246 & 6.2e-02 & 11.04 & 14.95 & 8.15 & 4.44 & -0.80 & $0.15\,\pm{0.05}$ \\
VAGC-411931 & 352.78012 & -1.21441 & 9.4e-02 & 10.99 & 15.90 & 7.05 & 5.43 & -0.66 & $0.43\,\pm{0.02}$ \\
GAMA-18375 & 36.66859 & -4.38636 & 7.7e-02 & 10.97 & 15.52 & 12.74 & 3.18 & -1.26 & $0.64\,\pm{0.03}$ \\
\ldots &
\ldots &
\ldots &
\ldots &
\ldots &
\ldots &
\ldots &
\ldots &
\ldots &
\ldots \\
GAMA-628 & 34.25908 & -5.06413 & 2.6e-02 & 8.95 & 16.90 & 1.78 & 3.53 & -1.07 & $0.17\,\pm{0.06}$ \\
GAMA-7807 & 35.85556 & -5.64420 & 1.9e-02 & 8.43 & 17.67 & 1.43 & 2.29 & - & $0.39\,\pm{0.05}$ \\
VAGC-96050 & 150.24617 & +3.46430 & 6.6e-03 & 8.34 & 15.95 & 2.85 & 2.79 & - & $0.44\,\pm{0.02}$ \\
\hline
\multicolumn{10}{c}{\textbf{Galaxies without stellar halo detection}} \\
\hline
VAGC-411901 & 350.41335 & -1.15126 & 7.6e-02 & 10.80 & 15.87 & 23.21 & 2.11 & -2.06 & $0.80\,\pm{0.01}$ \\
VAGC-527160 & 353.51266 & +0.36529 & 5.8e-02 & 10.80 & 15.35 & 4.01 & 4.95 & -0.39 & $0.69\,\pm{0.03}$ \\
VAGC-57421 & 148.66273 & +1.80068 & 9.9e-02 & 10.77 & 16.74 & 5.56 & 5.04 & -0.79 & $0.68\,\pm{0.05}$ \\
\ldots &
\ldots &
\ldots &
\ldots &
\ldots &
\ldots &
\ldots &
\ldots &
\ldots &
\ldots \\
VAGC-166177 & 352.75864 & -0.13193 & 8.5e-03 & 8.02 & 16.88 & 1.51 & 2.33 & - & $0.71\,\pm{0.03}$ \\
VAGC-57424 & 148.68743 & +1.60949 & 6.5e-03 & 7.78 & 16.92 & 1.13 & 2.61 & - & $0.60\,\pm{0.02}$ \\
VAGC-60459 & 151.10460 & +2.55873 & 3.8e-03 & 7.33 & 16.89 & 0.78 & 3.11 & - & $0.63\,\pm{0.02}$ \\
\enddata

\tablecomments{This table summarizes the basic properties of our high-inclination central galaxies, 
grouped according to whether a stellar halo is detected and sorted by galaxy stellar mass. 
Columns: (1) galaxy name; (2) RA (J2000); (3) Decl (J2000); (4) redshift; (5) galaxy stellar mass; 
(6) $r$-band magnitude; (7) half light radius along major axis; (8) concentration; (9) bulge fraction; 
(10) outer ellipticity. All properties are measured in imaging data corrected for PSF-induced scatter light. 
All colors and mass are $K$-corrected and Galactic-extinction-corrected. 
The uncertainties represent the $1\sigma$ confidence level.}
\tablecomments{Table \ref{tab:sampletbl_basic} is published in its entirety in the machine-readable format. 
A portion is shown here for guidance regarding its form and content. See the published version for the complete table.}
\end{deluxetable*}
\centerwidetable
\begin{deluxetable*}{l r r r r r r r r r}
\setlength{\tabcolsep}{2pt}
\tabletypesize{\small}
\tablecaption{Stellar halo properties of high-inclination central galaxies}
\label{tab:sampletbl_SH}
\tablehead{
\colhead{Name} &
\colhead{$(g-r)_{\rm halo}$} &
\colhead{$(g-i)_{\rm halo}$} &
\colhead{$(r-i)_{\rm halo}$} &
\colhead{$\log{M/L}_{\rm outer}$} &
\colhead{$n_{\rm g}$} &
\colhead{$n_{\rm r}$} &
\colhead{$n_{\rm i}$} &
\colhead{$\log(f_{\rm SH, mass})$} &
\colhead{$\log(f_{\rm SH, flux})$} \\
\colhead{} & \colhead{} & \colhead{} & \colhead{} & \colhead{[\msun/\Lsun]} & \colhead{} & \colhead{} & \colhead{} & \colhead{} &
\colhead{} \\
\colhead{(1)} &
\colhead{(2)} &
\colhead{(3)} &
\colhead{(4)} &
\colhead{(5)} &
\colhead{(6)} &
\colhead{(7)} &
\colhead{(8)} &
\colhead{(9)} &
\colhead{(10)}
}
\startdata
\hline
\multicolumn{10}{c}{\textbf{Galaxies with stellar halo detection}} \\
\hline
VAGC-13467 & $0.55\,\pm{0.05}$ & $0.92\,\pm{0.03}$ & $0.37\,\pm{0.04}$ & $0.12\,\pm{0.16}$ & $2.56\,\pm{0.09}$ & $2.81\,\pm{0.09}$ & $2.65\,\pm{0.07}$ & $-0.88\,\pm{0.23}$ & $-0.71\,\pm{0.02}$ \\
VAGC-411931 & $0.69\,\pm{0.04}$ & $0.98\,\pm{0.04}$ & $0.29\,\pm{0.03}$ & $0.28\,\pm{0.16}$ & $1.74\,\pm{0.09}$ & $1.81\,\pm{0.07}$ & $1.76\,\pm{0.06}$ & $-0.58\,\pm{0.23}$ & $-0.59\,\pm{0.02}$ \\
GAMA-18375 & $0.77\,\pm{0.19}$ & - & - & $0.38\,\pm{0.28}$ & - & - & - & $-1.30\,\pm{0.34}$ & $-1.43\,\pm{0.11}$ \\
\ldots &
\ldots &
\ldots &
\ldots &
\ldots &
\ldots &
\ldots &
\ldots &
\ldots &
\ldots \\
GAMA-628 & $0.45\,\pm{0.10}$ & $0.58\,\pm{0.08}$ & $0.13\,\pm{0.05}$ & $-0.01\,\pm{0.19}$ & $2.95\,\pm{0.14}$ & $2.63\,\pm{0.13}$ & $2.73\,\pm{0.13}$ & $-0.83\,\pm{0.26}$ & $-1.08\,\pm{0.05}$ \\
GAMA-7807 & $0.54\,\pm{0.06}$ & $0.70\,\pm{0.06}$ & $0.16\,\pm{0.06}$ & $0.10\,\pm{0.17}$ & $3.51\,\pm{0.31}$ & $3.54\,\pm{0.41}$ & $3.47\,\pm{0.44}$ & $-1.12\,\pm{0.26}$ & $-1.45\,\pm{0.09}$ \\
VAGC-96050 & $0.47\,\pm{0.05}$ & $0.58\,\pm{0.09}$ & $0.11\,\pm{0.08}$ & $0.02\,\pm{0.16}$ & $2.64\,\pm{0.41}$ & $3.04\,\pm{0.30}$ & $1.37\,\pm{0.61}$ & $-0.38\,\pm{0.24}$ & $-0.47\,\pm{0.05}$ \\
\hline
\multicolumn{10}{c}{\textbf{Galaxies without stellar halo detection}} \\
\hline
VAGC-411901 & - & - & - & - & - & - & - & $\leq$\,-1.66 & $\leq$\,-1.60 \\
VAGC-527160 & - & - & - & - & - & - & - & $\leq$\,-2.01 & $\leq$\,-1.91 \\
VAGC-57421 & - & - & - & - & - & - & - & $\leq$\,-2.08 & $\leq$\,-1.92 \\
\ldots &
\ldots &
\ldots &
\ldots &
\ldots &
\ldots &
\ldots &
\ldots &
\ldots &
\ldots \\
VAGC-166177 & - & - & - & - & - & - & - & $\leq$\,-1.55 & $\leq$\,-1.75 \\
VAGC-57424 & - & - & - & - & - & - & - & $\leq$\,-1.84 & $\leq$\,-2.02 \\
VAGC-60459 & - & - & - & - & - & - & - & $\leq$\,-1.43 & $\leq$\,-1.55 \\
\enddata

\tablecomments{This table presents the stellar halo properties of our high-inclination central galaxies, 
grouped according to whether a stellar halo is detected and sorted by the galaxy stellar mass. 
Columns: (1) galaxy name; (2) outer $g-r$ color; (3) outer $g-i$ color; (4) outer $r-i$ color; 
(5) outer mass-to-light ratio; (6) power-law index in $g$-band; (7) power-law index in $r$-band; 
(8) power-law index in $i$-band; (9) stellar halo mass fraction; (10) stellar halo flux fraction. 
All properties are measured in imaging data corrected for PSF-induced scatter light. 
All colors, mass-to-light ratio and mass are $K$-corrected and Galactic-extinction-corrected. 
The uncertainties represent the $1\sigma$ confidence level.}
\tablecomments{Table \ref{tab:sampletbl_SH} is published in its entirety in the machine-readable format. 
A portion is shown here for guidance regarding its form and content. See the published version for the complete table.}

\end{deluxetable*}
\clearpage{}

\end{document}